\documentclass[11pt,a4paper]{article}
\usepackage{jcappub}

\usepackage{url}
\usepackage{lscape}
\usepackage[toc,page]{appendix}
\usepackage{verbatim}
\usepackage{amsmath}
\usepackage{graphicx}

\usepackage{etoolbox}
\makeatletter
\pretocmd{\NAT@citexnum}{\@ifnum{\NAT@ctype>\z@}{\let\NAT@hyper@\relax}{}}{}{}
\makeatother

\def\beq{\begin{equation}}
\def\eeq{\end{equation}}
\def\alwaysmath#1{{\ifmmode{#1}\else{$#1$}\fi}}
\def\he#1{\hbox{\alwaysmath{{}^{#1}}{\rm He}}}
\def\li#1{\hbox{\alwaysmath{{}^{#1}}{\rm Li}}}

\def\sun{${\,_\odot}$}
\def\10830{{He~I $\lambda$10830}}
\def\3889{{He~I $\lambda$3889}}

\newcommand\hii{H\,{\sc II}}
\newcommand\hei{He\,{\sc I}}
\newcommand\hi{H\,{\sc I}}

\title{
Improving Helium Abundance Determinations with Leo P as a Case Study
}

\author[a]{Erik Aver}
\author[b]{Danielle~A.~Berg}
\author[c,d,e]{Keith~A.~Olive}
\author[f]{Richard~W.~Pogge}
\author[g]{John~J.~Salzer}
\author[c,e]{Evan~D.~Skillman}

\affiliation[a]{Department of Physics, Gonzaga University, \\
502 E Boone Ave, Spokane, WA 99258}
\emailAdd{aver@gonzaga.edu}
\affiliation[b]{Astronomy Department, University of Texas at Austin,\\
 Austin, TX 78712}
\emailAdd{daberg@austin.utexas.edu}
\affiliation[c]{School of Physics and Astronomy, University of Minnesota, \\
116 Church St. SE, Minneapolis, MN 55455}
\affiliation[d]{William I. Fine Theoretical Physics Institute, University of Minnesota, \\
116 Church St. SE, Minneapolis, MN 55455}
\emailAdd{olive@umn.edu}
\affiliation[e]{Minnesota Institute for Astrophysics, University of Minnesota, \\
116 Church St. SE, Minneapolis, MN 55455}
\emailAdd{skill001@umn.edu}
\affiliation[f]{Department of Astronomy, The Ohio State University, \\
140 W 18th Ave., Columbus, OH 43210}
\emailAdd{pogge.1@osu.edu}
\affiliation[g]{Department of Astronomy, Indiana University, \\
727 East Third St., Bloomington, IN 47405}
\emailAdd{josalzer@indiana.edu}

\abstract{
Currently, the primordial helium abundance is best estimated through spectroscopic observations of \hii\ regions in metal-poor galaxies. However these determinations are limited by several systematic uncertainties which ultimately limit our ability to accurately ascertain the primordial abundance.  In this study, we improve the methodologies for solving for the reddening, the emission contributions from collisional excitation of the \hi\ atoms, the effects underlying absorption in the \hi\ and \hei\ emission lines, and the treatment of the blended \hi\ and \hei\ emission at $\lambda$3889 with the aim of lowering the systematic uncertainties in helium abundance determinations.  To apply these methods, we have obtained observations of the \10830\ emission line in the brightest \hii\ region in the extremely metal-poor (3$\%$ Z$_{\odot}$) galaxy Leo~P with the LUCI1 instrument on the LBT.  We combine this measurement with previous MODS/LBT observations to derive an improved helium abundance. In doing so, our present analysis results in a decrease in the uncertainty in the helium abundance of Leo~P by approximately 70\%.  This result is combined with data from other observations to estimate the primordial helium mass fraction, Y$_{p}$ $=$ 0.2453 $\pm$ 0.0034.
}


\arxivnumber{}

\begin{document}

\begin{flushright}UMN-TH-4001/20\\FTPI-MINN-20/32\\
October 2020\end{flushright}

\maketitle
\flushbottom

\section{Introduction}\label{intro}

\subsection{The Primordial Helium Abundance}

Standard big bang nucleosynthesis (SBBN)\citep{wssok,osw,fs,gary07,iocco,cfoy} remains one of the deepest probes to the early universe. It is also an important probe of 
physics beyond the standard model. New physics, which affects the idealized conditions of 
a radiation dominated thermal bath at a temperature of roughly 1 MeV, has the potential
of upsetting the agreement between theory and the observational determinations of the $\he4$ and deuterium abundances. For example, theories which include new particle degrees of freedom
lead to an increased rate of expansion during nucleosynthesis leaving less time 
for the conversion of neutrons to protons, and hence to an increased helium mass fraction. These new particle degrees of freedom are often scaled as contributions
to the number of very light neutrinos, $N_\nu$. The upper limits on $N_\nu$ from BBN rely heavily on accurate measurements of the helium abundance (though limits based on D/H are becoming competitive \citep{fields2020}). 

Although the determination of the fundamental parameters of the standard cosmological model including dark matter and dark energy, $\Lambda$CDM, by WMAP \citep{wmap, wmap2} and Planck \citep{planck15,planck18} are unparalleled, big bang nucleosynthesis and the observations of the light element abundances also offer an important cross-check, in particular on the determination of the baryon density.  The most recent Planck result for the baryon density, $\Omega_B h^2 = 0.02237 \pm 0.00015$ \citep{planck18}, corresponds to a baryon-to-photon ratio of $\eta = (6.12 \pm 0.04) \times 10^{-10}$.  Because the uncertainty in $\eta$ is now less than 1\%,  SBBN (defined with $N_\nu = 3$) is a parameter-free theory \citep{cfo2}, and relatively precise predictions of the primordial abundances of the light elements D, $^{3}$He, $^{4}$He, and $^{7}$Li are available \citep{cfo,cfo3,coc2,cyburt,cuoco,serp,cfo5,pis,coc18,cfoy,fields2020}.  While the \li7 abundance remains problematic \citep{cfo5}, recent D/H determinations from quasar absorption systems have become quite precise, in their own right, and are in excellent agreement with the prediction from SBBN and the cosmic microwave background (CMB) \citep{dh,riemer,bala,cookeN,riemer17,zava,CPS}. Using a neutron mean life of $879.4 \pm 0.6$ s \citep{rpp}, SBBN yields a primordial abundance for $^{4}$He, $Y_{p}$, of $Y_p = 0.2469 \pm 0.0002$, using the Planck determined value of $\eta$ \citep{fields2020}.  By allowing $Y_{p}$ to vary as an independent parameter, fits to CMB anisotropies allow for a determination of $Y_p$ within the context of $\Lambda$CDM.  The recent Planck results found $Y_p = 0.239 \pm 0.013$ (68\% CL) \cite{planck18}.  Fortunately, the helium abundance from emission line measurements provide significantly better precision.

To test SBBN beyond D/H, it is clear that precise determinations of \he4 are necessary.
While it is unlikely that \he4 abundance measurements will ever be competitive with
the CMB and/or D/H determinations in fixing the baryon density, 
 \he4 provides the strongest constraints available on the physics of the early universe beyond the standard model \citep{cfos}.  Specifically, the measurement of $Y_{p}$ provides a constraint on the number of neutrino families, $N_\nu$.  Most combinations of 
 CMB data with abundance measurements used in BBN have a maximum likelihood
 with $N_\nu \approx 2.8$, just below the standard model value of 3; the uncertainty
 in this value depends sensitively on the data considered \citep{fields2020}. 
 For example, the CMB alone can fix $N_\nu$ only with a precision of $\pm 0.29$ (68\% CL).
 However, when Planck data \citep{planck18} are combined with the BBN relation
 between $Y_p$ and $\eta$, this uncertainty improves to $\pm 0.19$ without using 
 any abundance data. Helium abundance data further improves the uncertainty to $\pm 0.16$
 while D/H (without $Y_p$) gives $\pm 0.18$ and is competitive with \he4. Indeed,
 future CMB missions, such as CMB S4 \citep{CMB-S4,CMB-S4-RefDes}, are expected to improve these uncertainties $< \pm 0.09$ for the CMB alone, and $< \pm 0.06$ when the CMB is combined with BBN and $Y_p$ determinations. A well defined motivation of reducing this
 uncertainty is the ability to probe the effective number of degrees of freedom as seen by the CMB; $N_{\rm eff} = 3.045$ in the Standard Model \citep{pastor}. 
 
Obtaining better than 1\% precision for individual objects remains a challenge.  \he4 abundance determinations are generally fraught with systematic uncertainties and degeneracies among the input parameters needed to model emission line fluxes \citep{os01,os04,plp07,its07,AOS}. The primordial abundance of \he4 is determined by fitting the helium abundance versus a measurement of metallicity (e.g., oxygen), and extrapolating back to very low metallicity \citep{ptp74}.  Using Monte Carlo methods helps put these determinations on a firm statistical basis and allows an objective comparison between data and theoretical models \citep{AOS2,AOS3}. 
Previous studies \citep{AOS3,AOPS,AOS4} have found that the models used to extract abundances are statistically inconsistent with most of the available observations \citep{itl94,itl97,it98,it04,its07,it10,itg14} (in that those solutions fail to pass a standard 95\% CL $\chi^2$ test) and have employed corresponding quality cuts on their dataset \citep{AOS3}.  Recently, other analyses have found similar inconsistencies \citep{ftdt2,hcpb}, and others have employed stringent cuts on their observational data for related reliability concerns \citep{isg13, itg14}.
Whether these incompatibilities are due to deficiencies in
the model, or the data, or both is unknown. 

In this work, we address both fronts.
We discuss improvements to the modeling of emission line data and present
new Large Binocular Telescope (LBT) observations of the extremely metal-poor galaxy Leo~P.  For the optical \& near-IR spectrum, from the UV atmospheric cutoff to 1 $\mu$m, the LBT's Multi-Object Double Spectrograph \citep[LBT/MODS,][]{pogge2010} was used.  The LBT Utility Camera in the Infrared \citep[LUCI,][]{seifert2003} was used to obtain a near-IR spectrum, from 0.95 to 1.35 $\mu$m.  

There have been some recent model improvements which include newly calculated \he4 emissivities \citep{pfsd,pfsdc} and the addition of a near infrared line, \10830, \citep{itg14}. These have both
led to significant improvements in the \he4 abundance determinations \citep{AOPS,AOS4}.
The near infrared line is particularly useful as it helps break some of the degeneracies in the parameters used to derive the \he4 abundance, most notably that between the electron density, $n_e$ and temperature. 
It was found that the inclusion of this line not only reduced the uncertainty in the abundance determinations of individual objects, but also reduced the uncertainty in the primordial abundance \citep{AOS4}.
This resulted in a value for the primordial helium mass fraction, $Y_p = 0.2449 \pm 0.0040$ with slope $\Delta Y/\Delta {\rm (O/H)} = (78.9 \pm 43.3) {\rm O/H}$. Prior to the inclusion of \10830, the uncertainty in $Y_p$, was rather large, $\pm 0.0097$ \cite{AOPS}. 

We note also that there are several complimentary recent analyses of \he4.
In \citet{ppl}, five objects were used with an assumed slope of $\Delta Y/\Delta {\rm O} = 3.3 \pm 0.7$. They found $Y_p = 0.2446 \pm 0.0029$, where we presume that the smaller uncertainty in $Y_p$ is related to the smaller uncertainty assumed in $\Delta Y/\Delta \rm O$. 
In a similar approach, \citet{vpps} combine a very precise measurement of the He/H in the Small Magellanic Cloud H~II region NGC~346 and combine that with the same assumed slope of $\Delta Y/\Delta {\rm O}$ to obtain a value of $Y_p = 0.2451 \pm 0.0026$.
The recent analysis of \citet{ftdt}, using SDSS-III data with
regressions in O/H, N/H, and S/H, found $Y_p = 0.245 \pm 0.007$ which improved to
$Y_p = 0.243 \pm 0.005$, when Bayesian statistical techniques were applied \citep{ftdt2}. \citet{ftdt} use the collisionally excited [S~III] lines to derive a nebular temperature, and this appears to be a good choice for H~II regions where the quality of the spectra does not allow a measurement of the temperature from the He~I lines.
It is encouraging that all three of these independent analyses agree well within the limits of their uncertainties.

Most recently, new Keck NIRSPEC and Keck NIRES observations of \10830 in sixteen galaxies identified from SDSS imaging \citep{hsyu2018},
with existing optical spectra, were observed and analysed by \citet{hcpb} to obtain accurate \he4 abundances.
Their analysis followed the approach in \citep{AOS, AOS2,AOS3,AOPS,AOS4} and
they combined their new observations with a large number of observations from the literature.  From the combined dataset they obtained
$Y_p = 0.2436^{+0.0039}_{-0.0040}$ in remarkably close agreement with \citet{AOS4}. 
Importantly, \citet{hcpb} highlight the fact that only a small fraction of their newly observed targets passed their quality cuts for successful modeling.

\subsection{Potential Improvements for Determining Nebular Helium Abundances}

In \citet{AOS3}, we started with a dataset of 93 observations of 86 HII regions in 77 galaxies from \citet{its07}. It was argued in \citet{AOS3}, that the absence of He~I $\lambda$4026
in 23 of the observations led to a systematic bias towards higher \he4, and those observations without $\lambda$4026 were discarded. However, analysis of the remaining 
70 observations, found that in 45 of them, the comparison between model and data resulted in $\chi^2 > 4$, thus indicating a poor model fit to the data
at the 95.45\% CL, given nine observed line ratios and eight model parameters.
Further cuts were made due to solutions with very high ($> 25\%$) neutral hydrogen 
fraction, or very high oxygen abundances, leaving 22 observations of which
eight were flagged with either a large optical depth or large underlying absorption in either H or He. Thus, the majority of the observations were deemed unsuitable for further analysis. Using improved emissivities, two objects were rescued giving a total of 16 observations to deduce primordial \he4 \citep{AOPS}. 
A similar sized dataset (15 observations) was used in \citet{AOS4}, which incorporated \10830 observations from \citet{itg14}. Of the 93 observations from \citet{its07}, only 31 included He~I $\lambda$4026 and also had a \10830 observation in \citet{itg14}.  Despite the fact that not every object in our previous datasets \citep{AOS3, AOPS} had an observation for \10830, the additional observation softens the $\chi^2$ cut to $\chi^2 < 6$, due to the extra degree of freedom, and allows several previously excluded observations to be retained, resulting in a similarly sized dataset of 15 observations (see \citet{AOS4} for details). 

As noted earlier, the poor fit between the model and the data may be due to inherent
deficiencies in the model, or problems in the data. Here we consider improvements in 
both theory and observations. 
We utilize the LBT's higher resolution, higher sensitivity, and broader wavelength coverage spectrum. The large wavelength baseline enables us to use additional hydrogen and helium emission lines, which better constrain model parameters, particularly the reddening.
We improve our treatment of the collisional excitation of H~I and of underlying stellar absorption, as well as expand our modeling of both to incorporate the added emission lines.  We also revisit the procedure for handling the blended line \3889 with H8, including its underlying absorption and radiative transfer.  
As we describe in detail below, our model improvements come at the cost of 
adding one additional parameter, the underlying absorption for the hydrogen Paschen lines. However, we extend the calculation to include a net of eleven additional
flux ratios, thus making the model significantly more predicative. Of course,
even a perfect model would rely on high quality data in order to make accurate determinations of the model parameters. 
Improvements to the data can be addressed by increasing the sample with high quality observations. Here we make a first attempt by applying the model improvements to 
recent observations of Leo P. 

\subsection{Leo~P}

Leo P was discovered in the wide-field, unbiased \hi\ 21 cm ALFALFA survey \citep{giovanelli2013}.
It is a relatively isolated galaxy.
From HST observations of its RR Lyrae stars and the tip of its red giant branch, its distance has been determined to be 1.62 $\pm$ 0.15 Mpc \citep{mcquinn2013, mcquinn2015a} making
it a member of the NGC~3109 association of dwarf galaxies which is just outside of the Local Group zero-velocity surface \citep{tully2006}.
The stellar populations of Leo~P are consistent with relatively constant star formation over the 
Hubble time \citep{mcquinn2015a}. 
A bright H~II region was identified in H$\alpha$ imaging \citep{rhode2013}, allowing for chemical abundance analysis.  Recent, deeper, H$\alpha$ imaging has revealed more low surface brightness nebular emission associated with recent star formation \citep{evans2019}.

The oxygen abundance in Leo~P was first measured to be 12 + log(O/H) = 7.17 $\pm$ 0.04
\citep{skillman2013}, which is equivalent to 3\% of the solar value \citep{asplund2009}.
In part, because Leo~P lies on the metallicity-luminosity relationship defined by
nearby galaxies \citep{berg2012},
the low oxygen abundance is thought to be due to a combination of a low astration rate
and a low retention rate of the oxygen which is produced preferentially in supernovae
\citep{mcquinn2015b}.
This is in contrast to a large fraction of the known extremely metal-poor galaxies which 
appear to have had their oxygen abundance diluted by the recent infall of nearly pristine 
gas \citep[see discussions in][]{sanchez2017, mcquinn2020}.
\citet{skillman2013} measured a helium mass fraction in Leo~P of 
 0.2509$^{+0.0184}_{-0.0123}$.

\bigskip

This paper is organized as follows.
First, in section \ref{Model}, we discuss the basic theoretical model for the analysis of \he4 data, though details are saved for appendices \ref{App:Model} and \ref{App:Data}. In particular we outline the improvements we make to the model. 
As noted above, these include improvements to the treatment of underlying H and He absorption (with details in appendix \ref{App:UA}), the collisional correction to account for neutral H (details in appendix \ref{App:NHCC}), and the treatment of the blended He emission line $\lambda$3889 with H8. In
section \ref{Obs}, we provide new observational data from Leo P. 
An analysis of the new data including the improvements discussed in section \ref{Model}, is made in section \ref{Results} where we compute the \he4 abundance in Leo P and combine this result with the previous analysis made in \citet{AOS4}. 
Finally, section \ref{Conclusion} offers a discussion of the results and further prospects for improvement.

\section{Model Improvements} \label{Model}

We determine the \he4 abundance in an individual H~II region based on a Markov Chain Monte Carlo (MCMC) analysis. The MCMC method is an algorithmic procedure for sampling from a statistical distribution \citep{mar,met}.  The likelihood is simply, 
\beq
\mathcal{L}=\exp(-\chi^2/2),
\label{like}
\eeq
with $\chi^{2}$ given by, 
\beq
\chi^2 = \sum_{\lambda} \frac{(\frac{F(\lambda)}{F(H\beta|P\gamma)} - {\frac{F(\lambda)}{F(H\beta|P\gamma)}}_{meas})^2} { \sigma(\lambda)^2},
\label{eq:X2}
\eeq
where the emission line fluxes, $F(\lambda)$, are measured and calculated for a set of H and He lines, and $\sigma(\lambda)$ is the measured uncertainty in the flux ratio at each wavelength. 
The optical/near-IR emission line fluxes from the LBT/MODS spectrum are calculated relative to H$\beta$, while the infrared fluxes from the LBT/LUCI1 spectrum are calculated relative to the IR Paschen line, P$\gamma$.  Thus, by $F(H\beta|P\gamma)$, we mean $F(H\beta)$ for all lines other than \10830, and $F(P\gamma)$ for the latter. 

The base model, employed in \citet{AOS4}, made use of seven helium emission lines: $\lambda\lambda$3889, 4026, 4471, 5876, 6678, 7065, and 10830 in ratios to H$\beta$, as well as three hydrogen emission lines:
H$\alpha$, H$\gamma$, and H$\delta$ in ratios to H$\beta$.  In \citet{AOS4}, the \10830 flux was measured relative to $P\gamma$, but then scaled to calculate the \10830 flux relative to H$\beta$.  That scaling is no longer required in our updated model and approach (see appendix \ref{App:Model} for further details).  

These observed line ratios are used to fit eight parameters: the electron density, $n_e$, the electron temperature, $T_e$, the optical depth, $\tau$, the reddening coefficient, $C(H\beta)$, underlying absorption in H and He, $a_H$ and $a_{He}$, the neutral hydrogen fraction, $\xi$, and, of course, the helium abundance, $y^+$, of ionized He by number relative to H. Thus, we were left with two degrees of freedom for 10 observables. 

Empowered by the higher spectral resolution and broad wavelength coverage afforded by the LBT, we have invested in substantially extending and improving the theoretical model used to derive the \he4 abundance.  
Here we propose including two additional He line ratios, $\lambda$4922 and $\lambda$5015 relative to H$\beta$, as well as ten additional hydrogen line ratios:  H8, H9, H10, H11, H12, P8, P9, P10, P11, and P12, each relative to H$\beta$.  H7 is a blended line (with [Ne~III] $\lambda$3968) and is not considered.  Note that, because \3889 and H8 are blended, the inclusion of H8 does not represent a new observation, and \3889 + H8 is now incorporated as a blended line instead of attempting to deblend H8 from \3889 (see \S \ref{3889}). Along with the additional line ratios, we introduce one new parameter for the underlying absorption in the Paschen line, $a_P$ (see \S \ref{UA}). Thus, we are left with 12 degrees of freedom, corresponding to a total of 21 observed line ratios and 9 parameters to fit. 

The MCMC scans of our 9-dimensional parameter space map out the likelihood function and $\chi^2$ given above.  
We conduct a frequentist analysis, and the $\chi^2$ is minimized to determine the best-fit solution for the nine physical parameters, including $y^+$, as well as determining the ``goodness-of-fit''.  Uncertainties in each quantity are estimated by calculating a 1D marginalized likelihood and finding the 68\% confidence interval from the increase in the $\chi^2$ from the minimum.  
Expressions for the calculated flux ratios are given in appendix \ref{App:Model}. 

The increased number of emission lines extends the baseline over which reddening corrections are applied from $\Delta \lambda = $ 3039~\AA\ to 5775~\AA.  Further augmented by the increased number of emission lines employed, this greater baseline increases the sensitivity on the reddening parameter and correspondingly strengthens the determination of the reddening. 

Incorporating additional lines into our model requires the relevant atomic data and model coefficients, and improved data sources were used to obtain the required modelling values.  In particular, newer, improved data sources were employed to expand and improve the accuracy of our modelling of underlying stellar absorption and neutral hydrogen collisional excitation.  Furthermore, we revisited our approach to handling the blended line H8 + \3889.  

Our entire model and the data employed are detailed in the appendices (\ref{App:Model} \& \ref{App:Data}).  Here we will briefly introduce the three main areas of improvement in our modelling:  underlying stellar absorption, neutral hydrogen collisional excitation, and treatment of H8 + \3889.  

\subsection{Underlying Stellar Absorption} \label{UA}

The stellar continuum juxtaposes absorption features under nebular emission lines, and the emission lines in our model incorporate this effect \citep{os01, AOS}.  The wavelength-dependent underlying absorption in terms of equivalent width for the Balmer lines is scaled relative to H$\beta$, the Paschen lines are scaled relative to P$\gamma$, and He~I $\lambda$4471 is used for the helium lines.  For a given line, its scaling coefficient then multiplies the corresponding physical model parameter---a$_{H}$, a$_{P}$, or a$_{He}$---in determining that line's underlying absorption.  

As introduced in \citet{AOS}, the scaling coefficients employed for each line, relative to the underlying absorption of the chosen reference line, were calculated by F. Rosales-Ortega (private communication) using the Violent Star Formation Legacy tool ``SED@'' and using input models from \citet{gcmlh05} and \citet{mglc05}.  That work and those input models did not include all of the higher order lines added to our model in this work.  

The underlying absorption scaling coefficients were updated using the more advanced and inclusive spectral energy density output from the Binary Population and Spectral Synthesis (BPASS) stellar evolution and spectral synthesis models by Eldridge \& Stanway \citep{eld09, eld17, eld18}.  In particular, BPASS accounts for the significant effects of including binary star systems on stellar evolution and populations.  Absorption line equivalent widths, relative to the chosen reference line, were measured for the  hydrogen (Balmer and Paschen) and helium emission lines employed in our model for ages of 1, 2, 3 (3.16), 4, and 5 Myr.  The absorption line equivalent width ratios are relatively constant over the selected age range, as expected, and, correspondingly, the average value was adopted.  For the lines in common, the adopted scaling coefficients are consistent with those that were previously employed, based on the calculations by F. Rosales-Ortega.  

It should be noted that the ratios of the Paschen absorption lines to H$\beta$ are not relatively constant over the age range. However, they are relatively constant when calculated relative to P$\gamma$.  For this reason, we introduce another model parameter, a$_{P}$, for the underlying absorption for the Paschen lines (rather than also scaling their underlying absorption from a$_{H}$, as is done for the Balmer lines).  Appendix \ref{App:UA} provides further details and the adopted values.  

\subsection{Neutral Hydrogen Collisional Excitation} \label{NHCC}

\citet{ppl02} drew attention to the fact that the small amount of neutral hydrogen in the H~II region contributes to the emitted flux of the Balmer and Paschen lines through collisional excitation and that this affects the He/H abundance determination.  As a result, our model includes emission due to neutral hydrogen collisional excitation in its flux calculations.  

Introduced in \citet{AOS}, the collisional excitation rates were based on the effective collision strengths, $\Upsilon$, reported in \citet{and02} up to a principle quantum number of $n = 5$.  C.P. Ballance (private communication) graciously extended those calculations to $n = 8$.  These improved and extended collision strengths were used to update the collisional excitation rates.  

Our previous analyses only employed hydrogen emission lines up through H$\delta$, $n = 6$.  With the previously available collisional excitation data only going up through $n = 5$, the collision excitation for H$\delta$ ($n=6$) was estimated from H$\gamma$ ($n=5$) using a rough energy level scaling (see \citet{AOS} for further details).  With collision excitation data updated and expanded through $n = 8$, H$\delta$ is now directly calculated.  When we compare our previous collisional excitation rates for the four hydrogen emission lines ($n=3-6$) in our previous analysis with our new results, we find that H$\delta$ is the mostly strongly affected.  The collisional excitation rate for H$\delta$ calculated here (as given in appendix \ref{App:NHCC}) decreased compared to the estimate used in our previous work (as given in \citet{AOS}).  This effect is attributable to being able to directly calculate the collisional excitation rate from the new, expanded collisional strength data, rather than the rough scaling previously employed.  Next, we discuss a better approach to extrapolating for $n>8$.  

For H9, H10, H11, H12, P9, P10, P11 \& P12, collisional excitations to $n$ = 9, 10, 11, \& 12 are required.  As discussed in \citet{fmdzb16}, the effective collisional strength scales inversely proportional to the principle quantum number cubed, 
\beq
\Upsilon(n) \sim \frac{A}{(n+\alpha)^3}. 
\label{Upsilon}
\eeq
To estimate the collisional excitation for H9 \& P9, H10 \& P10, H11 \& P11, and H12 \& P12, the $\ell$ contributions for each level up to $n=8$ were summed and the $1/n^3$ scaling law above fit to extrapolate to $n$ = 9, 10, 11, \& 12.  Appendix \ref{App:NHCC} provides further details and the calculated coefficients for the collisional-to-recombination rate equations.    

\subsection{The Blended Emission Line H8 + \3889} \label{3889}

In our previous work, the helium emission line \3889 was incorporated into our model by subtracting the theoretical flux expected for H8 from the measured flux for the blended emission line H8 + \3889.  The theoretical flux expectation for H8 was calculated as a function of temperature, based on the data of \citet{hs87}, and updated as the temperature model parameter varied.  Corrections were made for underlying hydrogen absorption and for reddening.  See \citet{AOS} for further details and the parameterization.  

That approach is not completely self-consistent and suffers in its treatment of the model corrections for helium and hydrogen emission lines.  It can be improved by incorporating the observation into the model as a blended line, rather than attempting to deblend its contributions.  In this way, the sum of the theoretical fluxes for H8 \& \3889, each calculated using the relevant hydrogen and helium emission models, is compared to the observed blended emission line flux.  See appendix \ref{App:Model} and, in particular, eq.\:(\ref{eq:F_3889H8_EW}).  
Note also that, because H8 is only separated from \3889 by 0.4~\AA, and their Doppler broadening is approximately $\sim$0.4~\AA, H8, unlike any other hydrogen emission lines, also needs to be corrected for radiative transfer, $f_\tau(\lambda)$, just as \3889 is.  This additional term is reflected in eq.\:(\ref{eq:F_3889H8_EW}), and the helium and hydrogen contributions can be compared to eqs.\:(\ref{eq:F_He_EW}) \& (\ref{eq:F_H_EW}), respectively.

\section{New Observations} \label{Obs}

\subsection{Re-Analysis of the LBT/MODS Spectrum}

Optical spectroscopic observations of the bright H~II region in Leo~P were reported in \citet{skillman2013}.  The analysis there was based on a LBT observation taken with the MODS spectrograph \citep{pogge2010} on 2012 April 29.  Details of that observation are reported in \citet{skillman2013}, and we provide a brief summary here.
A total exposure time of 45 minutes was obtained with the MODS1 spectrograph, which provides continuous wavelength coverage from the ultraviolet atmospheric cutoff to 1 $\mu$m, with spectral resolutions of $\sim$2.4 \AA\ in the blue and $\sim$3.4 \AA\ in the red.

The LBT/MODS spectrum of Leo~P was one of the first spectra obtained with LBT/MODS and was reduced with the first version of the data analysis pipeline.  Since that time, several improvements have been made to the pipeline, most importantly, a more sophisticated treatment of the spectral sensitivity variations which arise due to the presence of the dichroic.  Thus, for the purposes of this new analysis, a complete re-reduction of the observations was performed with the new data reduction pipeline.  During this reduction we made a more optimal extraction in the spatial dimension (along the slit) which was narrow enough to optimize the signal/noise in the faintest emission lines but broad enough (20 0.12$^{\prime\prime}$ pixels for a total width of 2.4$^{\prime\prime}$) to insure a high-fidelity, robust measurement.

In table \ref{table:Fluxes}, we present the emission line fluxes, equivalent widths, and their associated uncertainties for all of the H and He recombination lines measured in the new analysis of the Leo~P LBT/MODS spectrum. 
As in \citet{skillman2013}, the uncertainties on the flux measurements were approximated using 
\begin{equation}
\sigma_{\lambda} \approx \sqrt{ {(2\times \sqrt{N}\times \zeta)}^2 + {(0.02\times F_{\lambda})}^2 },
\label{eq:uncertainty}
\end{equation}
where $N$ is the number of pixels spanning the Gaussian profile fit to the narrow emission lines and $\zeta$ is the rms noise in the continuum determined as the average of the rms measured on each side of an emission line.  For weak lines, the uncertainty is dominated by error from the continuum subtraction, so the rms noise term determines the uncertainty.  For the lines with flux measurements much stronger than the rms noise of the continuum, the error is dominated by the flux calibration.  A minimum flux uncertainty of 2\% was assumed, based on the uncertainties in the standard star measurements \citep{oke90}, and, for the strongest emission lines, it dominates the uncertainty estimate.

\subsection{New LBT LUCI1 NIR Observations}

As described in the introduction, measuring the \10830 emission line provides a strong constraint on the nebular density and breaks the temperature-density degeneracy, allowing for a reduced uncertainty in the He/H abundance.  For this reason, we have used the LBT's LUCI1 spectrograph \citep{seifert2003} to obtain a near-IR spectrum of the bright H~II region in Leo~P.  

\begin{figure}[ht!]
\resizebox{\textwidth}{!}{\includegraphics{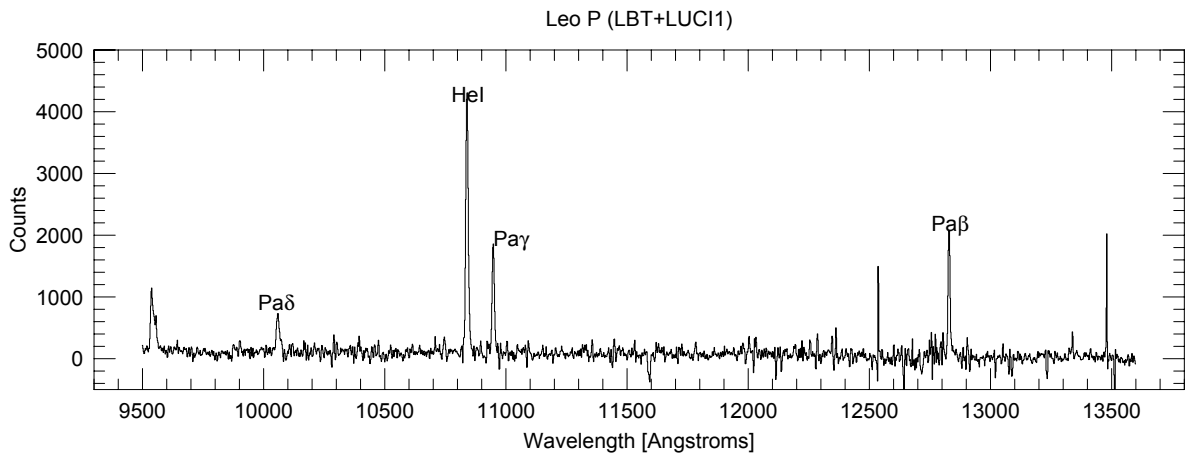}}
\caption{
Sky-subtracted, wavelength-calibrated near-infrared spectrum of Leo P obtained with the LUCI1 spectrograph on the Large Binocular Telescope.
}
\label{IRspec}
\end{figure}

A total integration time of two hours was obtained 2018 September 7.
The observations were 
taken with the facility 1$^{\prime\prime}$ wide long-slit mask (LS\_1.00arcsec), which is 205$^{\prime\prime}$ long, providing simultaneous measurement of the galaxy and surrounding night sky.  We took twelve 600-second integrations for a total of 2-hours on-sky, offsetting the target by $\pm$20$^{\prime\prime}$ along the slit between exposures to allow accurate subtraction of the bright terrestrial night-sky emission lines that dominate the raw spectra.
We used the low resolution 200H+K grating in the second order with the zJspec filter.
With this set-up, the NIR spectrum, shown in figure~\ref{IRspec}, covers the wavelength range from 0.95 to 1.35 $\mu$m 
at a spectral resolution of roughly $\lambda/\Delta\lambda\approx 2100$.
The wavelength calibration was determined using terrestrial OH airglow emission lines that appear in the long-slit spectra with the wavelengths tabulated by \citet{Rousselot2000}.

As shown in figure~\ref{IRspec}, the \10830 and Paschen $\gamma$ emission lines have both been detected at high significance.  Given their close proximity in wavelength, the flux ratio of these two lines, which is the measurement required for our determination of the He/H abundance ratio, did not require an observation of a flux standard.  The flux ratio for \10830 listed in table \ref{table:Fluxes} is the ratio of its emission line integrated counts relative to those of Paschen $\gamma$.  

\begin{table}[ht!]
\centering
\vskip .1in
\begin{tabular}{lcc}
\hline\hline
Ion 		        	& F($\lambda$)/F(H$\beta$)  &  W($\lambda$) [\AA]	\\	
\hline																								
H12 $\lambda$3750	&	0.0297	$\pm$	0.0034	&	2.55	\\
H11 $\lambda$3771	&	0.0346	$\pm$	0.0039	&	3.00	\\
H10 $\lambda$3798	&	0.0437	$\pm$	0.0041	&	3.77	\\
H9 $\lambda$3835	&	0.0580	$\pm$	0.0032	&	5.00	\\
He~I+H8 $\lambda$3889	&	0.1943	$\pm$	0.0060	&	17.09	\\
He~I $\lambda$4026	&	0.0121	$\pm$	0.0021	&	1.15	\\
H$\delta$ $\lambda$4101	&	0.2471	$\pm$	0.0074	&	25.29	\\
H$\gamma$ $\lambda$4340	&	0.4657	$\pm$	0.0134	&	56.53	\\
He I $\lambda$4471	&	0.0317	$\pm$	0.0024	&	4.25	\\
H$\beta$ $\lambda$4861	&	1.0000	$\pm$	0.0285	&	178	\\
He~I $\lambda$4922	&	0.0098	$\pm$	0.0018	&	1.77	\\
He~I $\lambda$5015	&	0.0242	$\pm$	0.0022	&	4.84	\\
He~I $\lambda$5876	&	0.1055	$\pm$	0.0036	&	35.60	\\
H$\alpha$ $\lambda$6563	&	2.9375	$\pm$	0.0835	&	1336	\\
He~I $\lambda$6678	&	0.0299	$\pm$	0.0012	&	14.43	\\
He~I $\lambda$7065	&	0.0252	$\pm$	0.0012	&	14.26	\\
P12 $\lambda$8750	&	0.0141	$\pm$	0.0022	&	19.61	\\
P11 $\lambda$8863	&	0.0150	$\pm$	0.0022	&	25.88	\\
P10 $\lambda$9015	&	0.0222	$\pm$	0.0019	&	37.87	\\
P9 $\lambda$9229	&	0.0270	$\pm$	0.0025	&	43.88	\\
P8 $\lambda$9546	&	0.0429	$\pm$	0.0048	&	95.59	\\
\hline	
 		        	& F($\lambda$)/F(P$\gamma$)  &  W($\lambda$) [\AA]	\\	
\hline	
He I $\lambda$10830	&	2.6617	$\pm$	0.0786	&	548.5	\\
\hline
\hline
\multicolumn{3}{l}{F(H$\beta$) = (2.991 $\pm$ 0.060) $\times$ 10$^{-15}$ erg s$^{-1}$ cm$^{-2}$} \\
\hline
\end{tabular}
\caption{Emission Line Fluxes and Equivalent Widths for Leo~P}
\label{table:Fluxes}
\end{table}

\section{Results} \label{Results}

\subsection{Leo~P}  \label{LeoP_Results}

Using our expanded and updated model, as outlined in \S \ref{Model} and detailed in Appendices \ref{App:Model} \& \ref{App:Data}, and the new observations and measurements discussed in \S \ref{Obs}, Leo~P was analyzed using our MCMC analysis (again see \S \ref{Model} \& appendix \ref{App:Model}).  The best-fit model parameter values and uncertainties are given in table \ref{table:LeoP}, with comparison to the previous analysis from \citet{skillman2013}.  The solutions agree well, though the uncertainties are dramatically reduced in our new analysis.  The total uncertainty range on y$^{+}$ decreased by just over 70\%, and most other parameters see even larger reductions in their uncertainty, especially $n_e$, $\tau$, $\xi$, and both $a_H$ and $a_{He}$. 

\begin{table}[ht!]
\centering
\vskip .1in
\begin{tabular}{lcc}
\hline\hline
                        & \citet{skillman2013}              &  This Work                    \\
\hline
Emission lines          &  9                                &  21                           \\
Free Parameters         &  8                                &  9                            \\
d.o.f.                  &  1                                &  12                           \\
95\% CL $\chi^2$        &  3.84                                &  21.03                           \\
\hline
He$^+$/H$^+$		    &  0.0837$^{+0.0084}_{-0.0062}$     &  0.0823$^{+0.0025}_{-0.0018}$ \\
n$_e$ [cm$^{-3}$]	    &  1$^{+206}_{-1}$                  &  39$^{+12}_{-12}$             \\
a$_{He}$ [\AA]		    &  0.50$^{+0.42}_{-0.42}$           &  0.42$^{+0.11}_{-0.15}$       \\
$\tau$				    &  0.00$^{+0.66}_{-0.00}$           &  0.00$^{+0.13}_{-0.00}$       \\
T$_e$ [K]			    &  17,060 $^{+1900}_{-2900}$        &  17,400 $^{+1200}_{-1400}$    \\
C(H$\beta$)			    &  0.10$^{+0.03}_{-0.07}$           &  0.10$^{+0.02}_{-0.02}$       \\
a$_H$ [\AA]			    &  0.94$^{+1.44}_{-0.94}$           &  0.51$^{+0.17}_{-0.18}$       \\
a$_P$ [\AA]             &  -                                &  0.00$^{+0.52}_{-0.00}$       \\
$\xi$ $\times$ 10$^4$   &  0$^{+156}_{-0}$                  &  0$^{+7}_{-0}$                \\
$\chi^2$			    &  3.3                              &  15.3                         \\
p-value                 &  7\%                              &  23\%                         \\
\hline
O/H $\times$ 10$^5$	    &  1.5 $\pm$ 0.1                    &  1.5 $\pm$ 0.1                \\
Y				        &  0.2509 $\pm$ 0.0184              &  0.2475 $\pm$ 0.0057          \\
\hline
\end{tabular}
\caption{Physical conditions, He$^+$/H$^+$ abundance solution, and regression values of Leo~P}
\label{table:LeoP}
\end{table}

The primary driver of the reduced uncertainty on the helium abundance determination is the addition of \10830, accounting for approximately half of the reduction.  This is in keeping with the results observed in \citet{itg14} and \citet{AOS4}.  Though none of the other added emission lines carry the same weight as \10830 in constraining the solution, together their impact is just as significant.  Most of the added emission lines (9 of 12) are hydrogen emission lines, and, correspondingly, their impact primarily results in much stronger constraints on the reddening, due to the significantly extended wavelength baseline, and the underlying hydrogen absorption.  The required addition of a model parameter, the underlying absorption for the Paschen lines, $a_P$, separate from that for the Balmer lines, $a_H$, reduces the gain slightly, but with five Paschen lines utilized (and the promise of two more in future LBT Optical + IR observations), $a_P$ is still well-constrained, and the Paschen lines contribute substantially to reducing the uncertainty (primarily through the reddening).  

Table \ref{table:chi} provides a full comparison of the measured fluxes and the model predictions for Leo~P broken down by line.  The relative uncertainty for each measured flux is also included.  The stronger lines dominate the $\chi^2$ minimization and best-fit solution, due to their lower relative uncertainties. However, the weaker lines, appropriately weighted by their larger uncertainties, still contribute to constrain the best-fit solution and parameter uncertainties.  

Our previous analysis of the Leo P spectrum yielded a fit to eight parameters
using 9 nine emission line ratios \citep{skillman2013}. As indicated in table \ref{table:LeoP}, the total $\chi^2$ for the best fit solution was 3.3 for a single degree of freedom,
corresponding to a p-value of 7\%. 
Though our new solution has a significantly higher value of 
$\chi^2 = 15.3$ (found by summing the $\chi^2$ contributions in the rightmost column in table \ref{table:chi} for the 21 emission line ratios), 
it corresponds to 12 degrees of freedom, and has a p-value of 23\%.

In addition to an overall improvement in the fit, 
we see from the last row of table \ref{table:LeoP}, a significant drop in the uncertainty in 
the resulting helium mass fraction. The new uncertainty is more than a factor of three
smaller than the previous result. The central value, in contrast, changed by less than 1 $\sigma$ leading to a helium mass fraction in Leo~P of  $Y = 0.2475 \pm 0.0057$. 

\begin{table}[ht!]
\centering
\vskip .1in
\begin{tabular}{lccccc}
\hline\hline
Emission Line 		        	& $(\frac{\textrm{F}(\lambda)}{\textrm{F(H}\beta)})_{\textrm{meas}}$  &  $\sigma$    & \%  & $(\frac{\textrm{F}(\lambda)}{\textrm{F(H}\beta)})_{\textrm{mod}}$   &   $\chi^2$    \\	
\hline																								
\multicolumn{6}{l}{\textit{Helium}:}    \\
He~I $\lambda$4026	&	0.0121	&	0.0021	&	17.30	&	0.0126	&		0.048	\\
He I $\lambda$4471	&	0.0317	&	0.0024	&	7.67	&	0.0343	&		1.171	\\
He~I $\lambda$4922	&	0.0098	&	0.0018	&	18.25	&	0.0089	&		0.290	\\
He~I $\lambda$5015	&	0.0242	&	0.0022	&	9.23	&	0.0245	&		0.014	\\
He~I $\lambda$5876	&	0.1055	&	0.0036	&	3.39	&	0.1046	&		0.064	\\
He~I $\lambda$6678	&	0.0299	&	0.0012	&	4.08	&	0.0301	&		0.036	\\
He~I $\lambda$7065	&	0.0252	&	0.0012	&	4.79	&	0.0258	&		0.272	\\
\multicolumn{6}{l}{\textit{Hydrogen}:}	\\
H12 $\lambda$3750	&	0.0297	&	0.0034	&	11.40	&	0.0245	&		2.351	\\
H11 $\lambda$3771	&	0.0346	&	0.0039	&	11.28	&	0.0331	&		0.139	\\
H10 $\lambda$3798	&	0.0437	&	0.0041	&	9.37	&	0.0456	&		0.230	\\
H9 $\lambda$3835	&	0.0580	&	0.0032	&	5.53	&	0.0647	&		4.388	\\
H$\delta$ $\lambda$4101	&	0.2471	&	0.0074	&	2.99	&	0.2465	&		0.007	\\
H$\gamma$ $\lambda$4340	&	0.4657	&	0.0134	&	2.88	&	0.4554	&	0.600	\\
H$\alpha$ $\lambda$6563	&	2.9375	&	0.0835	&	2.84	&	2.9614	&	0.082	\\
P12 $\lambda$8750	&	0.0141	&	0.0022	&	15.94	&	0.0114	&		1.422	\\
P11 $\lambda$8863	&	0.0150	&	0.0022	&	14.96	&	0.0149	&		0.004	\\
P10 $\lambda$9015	&	0.0222	&	0.0019	&	8.44	&	0.0199	&		1.501	\\
P9 $\lambda$9229	&	0.0270	&	0.0025	&	9.09	&	0.0276	&		0.048	\\
P8 $\lambda$9546	&	0.0429	&	0.0048	&	11.25	&	0.0397	&		0.430	\\
\multicolumn{6}{l}{\textit{Blended He}+\textit{H}:}	\\
He~I+H8 $\lambda$3889	&	0.1943	&	0.0060	&	3.08	&	0.1856		&	2.156	\\
\hline
 		        	& $(\frac{\textrm{F}(\lambda)}{\textrm{F(P}\gamma)})_{\textrm{meas}}$  &  $\sigma$    & \%  & $(\frac{\textrm{F}(\lambda)}{\textrm{F(P}\gamma)})_{\textrm{mod}}$   &   $\chi^2$    \\
\hline
He I $\lambda$10830	&	2.6617	&	0.0786	&	2.95	&	2.6578	&		0.002	\\
\hline
\end{tabular}
\caption{Comparison of the measured flux ratio to the model prediction by line for the best-fit solution for Leo~P. }
\label{table:chi}
\end{table}

\subsection{The Primordial Helium Abundance} \label{Yp}

A regression of Y, the helium mass fraction, versus O/H, the oxygen abundance, from nearby galaxies, is used to extrapolate to the primordial value\footnote{This work takes $Z=20(O/H)$ such that $Y=\frac{4y(1-20(O/H))}{1+4y}$}.  The O/H values are taken directly from \citet{its07}, except for Leo~P, where the value is taken from \citet{skillman2013}.  

\begin{figure}
\resizebox{\textwidth}{!}{\includegraphics{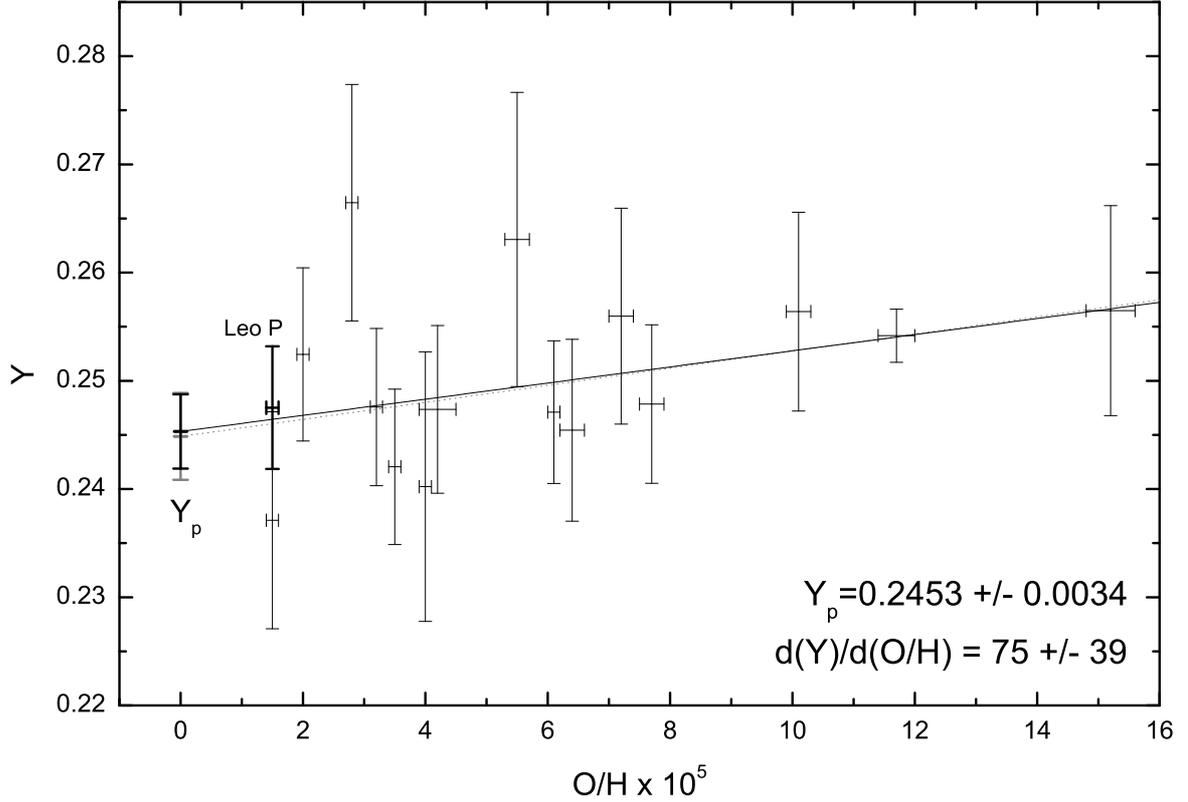}}
\caption{
Helium abundance (mass fraction) versus oxygen to hydrogen ratio regression calculating the primordial helium abundance.  The previous result from \citet{AOS4} is lighter grey, with the new result, based on including Leo~P, shown with a solid fit line and a bolded intercept.  The added point, Leo~P based on the analysis of this work, is also shown as bold. The light grey point at the same value of O/H as Leo~P is I~Zw~18. 
}
\label{Y_OH}
\end{figure}

The relevant values for the regression are given in table \ref{table:PH}.  The regression is based on the results from \citet{AOS4}, combined with the results for Leo~P from this work.  There are 15 objects in the qualifying dataset of \citet{AOS4}, and Leo~P is added to the previous results for those 15.  Regression based on those 16 points yields
\beq
Y_p = 0.2453 \pm 0.0034,
\label{eq:Yp}
\eeq
with a slope of 75 $\pm$ 39 and a total $\chi^2$ of 7.7.  The result is shown in figure~\ref{Y_OH}. This result for $Y_{p}$ agrees well with the SBBN value of $Y_p = 0.2469 \pm 0.0002$ \citep{fields2020}, based on the Planck determined baryon density \cite{planck18}.  Eq.\:(\ref{eq:Yp}) also agrees well with the SBBN-independent, direct Planck estimation of $Y_p = 0.239 \pm 0.013$ \citep{planck18}, using temperature, polarization, and lensing data, which is not surprising given the relatively large uncertainty on the CMB measurement. 

\citet{AOS4} determined $Y_p = 0.2449 \pm 0.0040$ with a slope of 79 $\pm$ 43.  Since the result given in eq.\:(\ref{eq:Yp}) is based on the same dataset, with only one additional point, Leo~P, it is not surprising that the results are so similar.  Perhaps more surprising is that the addition of a single point can affect the result and its uncertainty a noticeable amount.  Leo~P's impact on the regression comes from two main sources.  First, Leo~P's very low metallicity gives it greater weight in determining the intercept of the regression.  Second, Leo~P's uncertainty is the second lowest in the regression dataset.  

All of the results from \citet{AOS4} are based on observations which include \10830.  Leo~P's further reduced uncertainty compared to the typical level seen from \citet{AOS4} is primarily due to its expanded number of emission lines employed, as enabled by the LBT's higher-resolution and broader wavelength coverage spectra.  The change in the intercept is seen in figure~\ref{Y_OH},
where the new result is bold and the previous result from \citet{AOS4} (partially covered by the new result) is lighter grey.  The previous result does not include Leo P which is labelled on the figure and shown as bold (and which partially overlaps with I~Zwicky~18 at the same O/H).

\citet{hcpb} recommends a regression in terms of $y$ instead of $Y$, in part to eliminate the dependence on $Z$ (through $Y=\frac{4y(1-Z)}{1+4y}$) in determining $Y_p$.  Performing our regression in terms of $y$, and then converting from $y_p$ to $Y_p$, we obtain: 
\begin{align}
y_p & = 0.0811 \pm 0.0015  \\
Y_p & = 0.2450 \pm 0.0035.
\label{eq:yp}
\end{align}
As one can see, the regression in terms of $y$ leads to a result in excellent agreement with eq.\:(\ref{eq:Yp}). 

A comparison of our result with other recent determinations is found in table \ref{table:YpComp}.  It is encouraging that all of the most recent results agree.  The only one not in agreement with the others is \citet{itg14}. See \citet{AOS4} for a detailed description of the differences in the methodology and dataset cuts between \citet{itg14} and our model and approach.  

\begin{table}[ht!]
\small
\centering
\vskip .1in
\begin{tabular}{lcccc}
\hline\hline
Object & 	He$^+$/H$^+$ 	      & He$^{++}$/H$^+$     & Y 		  & O/H $\times$ 10$^5$ \\
\hline
Leo~P           &   0.08227  $\pm$  0.00250 &   0.0000	$\pm$	0.0000  &   0.2475  $\pm$   0.0057  & 1.5 $\pm$   0.1 \\
\hline
I~Zw~18~SE~1	&	0.07693	$\pm$	0.00423	&	0.0008	$\pm$	0.0008	&	0.2371	$\pm$	0.0100	&	1.5	$\pm$	0.1	\\
SBS~0335-052E1+3	&	0.08201	$\pm$	0.00303	&	0.0026	$\pm$	0.0018	&	0.2524	$\pm$	0.0080	&	2	$\pm$	0.1	\\
J0519+0007	&	0.08875	$\pm$	0.00461	&	0.0021	$\pm$	0.0021	&	0.2665	$\pm$	0.0109	&	2.8	$\pm$	0.1	\\
SBS~0940+544~2	&	0.08179	$\pm$	0.00316	&	0.0005	$\pm$	0.0005	&	0.2476	$\pm$	0.0073	&	3.2	$\pm$	0.1	\\
Tol~65	&	0.07883	$\pm$	0.00294	&	0.0011	$\pm$	0.0011	&	0.2421	$\pm$	0.0072	&	3.5	$\pm$	0.1	\\
SBS~1415+437~(No.~1)~3	&	0.07694	$\pm$	0.00494	&	0.0022	$\pm$	0.0022	&	0.2402	$\pm$	0.0125	&	4	$\pm$	0.1	\\
SBS~1415+437~(No.~2)	&	0.08226	$\pm$	0.00343	&	0.0000	$\pm$	0.0000	&	0.2474	$\pm$	0.0078	&	4.2	$\pm$	0.3	\\
CGCG~007-025~(No.~2)	&	0.08867	$\pm$	0.00623	&	0.0007	$\pm$	0.0007	&	0.2631	$\pm$	0.0136	&	5.5	$\pm$	0.2	\\
Mrk~209	&	0.08114	$\pm$	0.00272	&	0.0011	$\pm$	0.0011	&	0.2471	$\pm$	0.0066	&	6.1	$\pm$	0.1	\\
SBS~1030+583	&	0.07937	$\pm$	0.00306	&	0.0021	$\pm$	0.0021	&	0.2454	$\pm$	0.0084	&	6.4	$\pm$	0.2	\\
Mrk~71~(No.~1)	&	0.08539	$\pm$	0.00445	&	0.0008	$\pm$	0.0008	&	0.2560	$\pm$	0.0100	&	7.2	$\pm$	0.2	\\
SBS~1152+579	&	0.08139	$\pm$	0.00303	&	0.0012	$\pm$	0.0012	&	0.2478	$\pm$	0.0073	&	7.7	$\pm$	0.2	\\
Mrk~59	&	0.08548	$\pm$	0.00405	&	0.0010	$\pm$	0.0010	&	0.2564	$\pm$	0.0092	&	10.1	$\pm$	0.2	\\
SBS~1135+581	&	0.08462	$\pm$	0.00072	&	0.0008	$\pm$	0.0008	&	0.2542	$\pm$	0.0025	&	11.7	$\pm$	0.3	\\
Mrk~450~(No.~1)	&	0.08634	$\pm$	0.00441	&	0.0003	$\pm$	0.0003	&	0.2565	$\pm$	0.0097	&	15.2	$\pm$	0.4	\\
\hline
\end{tabular}
\caption{Primordial helium regression values}
\label{table:PH}
\end{table}

\begin{table}[ht!]
\centering
\vskip .1in
\begin{tabular}{llrr}
\hline\hline
Citation & Y$_\textrm{P}$ & N & Method \\
\hline
\citeauthor{itg14} (\citeyear{itg14}) \citep{itg14}                 &   0.2551 $\pm$ 0.0022     &   28      &   H~II Region     \\
\citeauthor{AOS4} (\citeyear{AOS4}) \citep{AOS4}                    &   0.2449 $\pm$ 0.0040     &   15      &   H~II Region     \\
\citeauthor{ppl} (\citeyear{ppl}) \citep{ppl}                       &   0.2446 $\pm$ 0.0029     &   5       &   H~II Region     \\
Planck Collaboration (\citeyear{planck18}) \citep{planck18}         &   0.239 $\pm$ 0.013   &   &   CMB     \\
\citeauthor{cooke18} (\citeyear{cooke18}) \citep{cooke18}           &   0.250$^{+0.033}_{-0.025}$   &   1   &   Absorption Line     \\
\citeauthor{vpps} (\citeyear{vpps}) \citep{vpps}                    &   0.2451 $\pm$ 0.0026     &   1       &   H~II Region     \\
\citeauthor{ftdt2} (\citeyear{ftdt2}) \citep{ftdt2}                 &   0.243 $\pm$ 0.005       &   16      &   H~II Region     \\
\citeauthor{hcpb} (\citeyear{hcpb}) \citep{hcpb}                    &   0.2436$^{+0.0039}_{-0.0040}$    &   54  &   H~II Region     \\
This Work       &   0.2453 $\pm$ 0.0034     &   16      &   H~II Region     \\
\hline
\citeauthor{fields2020} (\citeyear{fields2020}) \citep{fields2020}  &   0.2469 $\pm$ 0.0002 &           &   SBBN + CMB            \\
\hline
\end{tabular}
\caption{Recent Primordial Helium Abundance Results} 
\label{table:YpComp}
\end{table}

\section{Discussion} \label{Conclusion}

Fitting a large number of parameters, with a relatively small number of observations is 
a dangerous proposition. In this work, we have advocated for a serious expansion
in the number of observable emission lines used to fit our model parameters. 
We have shown that moving from eight line ratios (\10830 was not previously available for Leo P), to the present 21 at the expense of adding only a single model parameter (for underlying 
absorption in the hydrogen Paschen lines), leads to a significant improvement in the determinations of the physical parameters, including the helium abundance. 

However, adding to the number of observables requires a host of model improvements
that have been highlighted above, and described in detail in the appendices. 
Among these improvements is the wavelength dependence of underlying absorption. 
Here we use a more sophisticated set of scaling coefficients derived from BPASS stellar evolution and spectral synthesis models. Adding the higher principle quantum number lines (up to $n=12$)
requires a more detailed set of H collisional corrections. We use an expanded network
up to $n = 8$ and an analytical scaling given by eq.\:(\ref{Upsilon}) for higher values of $n$.
Finally, we advocate for a different treatment of the blended H8 + \3889
line. Rather than attempting to subtract out the H8 contribution and treat the observation
as a He emission line, we sum the theoretical fluxes of the two, and compare directly with the observation. 

Beyond our demonstrated model improvements, we are currently statistics limited. 
Clearly, enlarging the sample of extremely metal-poor (XMP) galaxies\footnote{Here we adopt the definition of an XMP galaxy as having 12 + log(O/H) $\le$ 7.3 \citep{mcquinn2020}.}, with well determined He abundances,
will lead to improved constraints on the value of Y$_p$.
The XMP galaxies are the most relevant for measurements constraining Y$_p$;
however, historically, the searches for these galaxies have had very low yields
(see discussion in \citet{sanchez2017}).  
Fortunately, in recent years, there have been a number of discoveries of XMP galaxies \citep{hirschauer2016,yang2017,guseva2017,hsyu2017,hsyu2018,izotov2018, izotov2019,senchyna2019,pustilnik2020,kojima2020}. 
Although many of these XMP galaxies are too low in emission line surface brightness to be
viable for precise He abundance measurements, future observations of the most promising 
of these galaxies, with quality comparable to the observations of Leo~P presented here,
represents a very desirable goal.

\acknowledgments

The authors owe special thanks to Connor Ballance and Martin O'Mullane for their effective collision strength calculations for electron impact excitation for neutral hydrogen.  
This work made use of v2.2 of the Binary Population and Spectral Synthesis (BPASS) models, as described in Eldridge, Stanway et al.\:(2017) and Stanway \& Eldridge et al.\:(2018).  
The work of KAO is supported in part by DOE grant DE-SC0011842.  EDS is grateful for partial support from the University of Minnesota.  
EA benefited greatly from three visits during sabbatical to the University of Minnesota and is grateful to the University of Minnesota and the William I. Fine Theoretical Physics Institute for the support.  

This paper uses data taken with the MODS spectrographs built with funding from NSF grant AST-9987045 and the NSF Telescope System Instrumentation Program (TSIP), with additional funds from the Ohio Board of Regents and the Ohio State University Office of Research.
This paper made use of the modsIDL spectral data reduction pipeline developed in part with funds provided by NSF Grant AST-1108693.
This work was based in part on observations made with the Large Binocular Telescope (LBT). The LBT is an international collaboration among institutions in the United States, Italy and Germany. The LBT Corporation partners are: the University of Arizona on behalf of the Arizona university system; the Istituto Nazionale di Astrofisica, Italy; the LBT Beteiligungsgesellschaft, Germany, representing the Max Planck Society, the Astrophysical Institute Potsdam, and Heidelberg University; the Ohio State University; and the Research Corporation, on behalf of the University of Notre Dame, the University of Minnesota, and the University of Virginia.
This research has made use of the NASA/IPAC Extragalactic Database (NED) which is operated by the
Jet Propulsion Laboratory, California Institute of Technology, under contract with the
National Aeronautics and Space Administration and the NASA Astrophysics Data System (ADS).

\begin{appendices}
 
Appendices \ref{App:Model} \& \ref{App:Data} detail our entire model, including the updates and improvements made as part of this work, and the data employed.  With the goal of providing a more cohesive and seamless presentation of our model, some content from \S \ref{Model} is reproduced.  

\section{H~II Region Helium Abundance Model} \label{App:Model}

Our method for determining the \he4 abundance in an individual H~II region is based on a Markov Chain Monte Carlo (MCMC) analysis. The MCMC method is an algorithmic procedure for sampling from a statistical distribution \citep{mar,met}.  The likelihood is, 
\beq
\mathcal{L}=\exp(-\chi^2/2),
\label{app:eq:L}
\eeq
with $\chi^{2}$ given by, 
\beq
\chi^2 = \sum_{\lambda} {(\frac{F(\lambda)}{F(H\beta|P\gamma)} - {\frac{F(\lambda)}{F(H\beta|P\gamma)}}_{meas})^2 \over \sigma(\lambda)^2},
\label{app:eq:X2}
\eeq
where the emission line fluxes, $F(\lambda)$, are measured and calculated for a set of H and He lines.  The optical/near-IR emission line fluxes from the LBT/MODS spectrum are calculated relative to H$\beta$, while the infrared fluxes from the LBT/LUCI spectrum are calculated relative to the IR Paschen line, P$\gamma$.  This difference in the reference line for each spectrum is encoded in eq.\:(\ref{app:eq:X2}) by $F(H\beta|P\gamma)$, where the appropriate reference line flux is chosen for the target emission line based on its source ($F(H\beta)$ for MODS Optical/NIR or $F(P\gamma)$ for LUCI IR).  The emission lines employed in this work, with their reference lines specified, are as follows.  

Eight helium line ratios are employed:  $\lambda\lambda$4026, 4471, 4922, 5015, 5876, 6678, and 7065, relative to H$\beta$, and $\lambda$10830, relative to P$\gamma$. Twelve hydrogen line ratios are employed:  H$\alpha$, H$\gamma$, H$\delta$, H9, H10, H11, H12, P8, P9, P10, P11, and P12, relative to H$\beta$.  Our expanded model incorporates P$\beta$ and P$\delta$, as well, but the lack of a flux calibration for the LUCI1 IR spectrum counseled against their use for this work (see \S \ref{Obs}).  Finally, the blended line \3889 + H8, relative to H$\beta$, is also employed.  

Our model for the calculated, theoretical fluxes depends on nine physical parameters associated with the H~II region: electron density, $n_e$; temperature, $T$; optical depth $\tau$; underlying stellar He and H absorption, including separate underlying absorption parameters for the Balmer and Paschen lines (see \S \ref{UA}), $a_{He}$, $a_H$, $a_P$; reddening, $C(H\beta$); the fraction of neutral hydrogen, $\xi$; and, of course, the He abundance, parameterized in terms of the abundance by number (relative to H) of ionized He, $y^+$.  We utilize MCMC scans of our 9-dimensional parameter space to map out the likelihood function and $\chi^2$ given above (eqs.\:(\ref{app:eq:L}) \& (\ref{app:eq:X2})).  

We conduct a frequentist analysis, and the $\chi^2$ is minimized to determine the best-fit solution for the nine physical parameters, as well as determining the ``goodness-of-fit''.  Uncertainties in each quantity are estimated by calculating a 1D marginalized likelihood and finding the 68\% confidence interval from the increase in the $\chi^2$ from the minimum ($\Delta\chi^2=1$).  

The flux ratios used in eq.\:(\ref{app:eq:X2}) are calculated using the following equations for the He and H lines.  
In the following equations for $F(\lambda)$, $W(\lambda)$ is the measured equivalent width.  The emissivity, $E$, including collisional corrections for the He lines, is a function of $n_e$ and $T$ and is discussed in more detail in appendix \ref{App:Emissivities}.  The term $10^{-f(\lambda)C(H\beta)}$ accounts for reddening.  The function $f_\tau(\lambda)$ represents a correction for radiative transfer and depends on $\tau$, $n_e$, and $T$.  Appendix \ref{App:RRT} gives our treatment of reddening and radiative transfer.  The underlying absorption corrections are wavelength-dependent, and, as mentioned above, are characterized by three parameters, $a_{He}$, $a_H$ and $a_P$, for the the helium, Balmer, and Paschen lines, respectively.  The underlying absorption correction for each line, $a(\lambda)$, is determined by its underlying absorption scaling coefficient times the relevant underlying absorption model parameter ($a_{He}$, $a_H$, or $a_P$), as given in appendix \ref{App:UA}.  The hydrogen collisional corrections are accounted for by $\frac{C}{R}$, which depends on the fraction of neutral-to-ionized hydrogen, $\xi$, and $T$, as detailed in appendix \ref{App:NHCC}.

The optical Helium line flux ratios are calculated relative to H$\beta$ and corrected for underlying absorption based upon $a_{He}$ and $a_H$.  
\beq
\frac{F(\lambda)}{F(H\beta)} = y^{+}\frac{E(\lambda)}{E(H\beta)}{\frac{W(H\beta)+a_{H}(H\beta)}{W(H\beta)} \over \frac{W(\lambda)+a_{He}(\lambda)}{W(\lambda)}}{f_{\tau}(\lambda)}\frac{1}{1+\frac{C}{R}(H\beta)}10^{-f(\lambda)C(H\beta)}
\label{eq:F_He_EW}
\eeq
The Balmer H flux ratios are calculated analogously, relative to H$\beta$, and corrected for underlying absorption based upon $a_H$.  
\beq
\frac{F(\lambda)}{F(H\beta)} = \frac{E(\lambda)}{E(H\beta)}{\frac{W(H\beta)+a_{H}(H\beta)}{W(H\beta)} \over \frac{W(\lambda)+a_{H}(\lambda)}{W(\lambda)}}\frac{1+\frac{C}{R}(\lambda)}{1+\frac{C}{R}(H\beta)}10^{-f(\lambda)C(H\beta)}
\label{eq:F_H_EW}
\eeq
The IR He line, \10830, is calculated relative to P$\gamma$, and P$\gamma$ is corrected for underlying absorption based on $a_P$, while \10830 is corrected for underlying absorption based on $a_{He}$.  Note that in AOS4 \citep{AOS4}, the measured \10830 flux ratio relative to P$\gamma$ was scaled using the theoretical ratio for F(P$\gamma$)/F(H$\beta$) to estimate F(\10830)/F(H$\beta$).  That F(\10830)/F(H$\beta$) value was then incorporated into our model analogously to all the other (optical) helium emission lines (which are all measured and input relative H$\beta$).  Here, we incorporate the measured F(\10830)/F(P$\gamma$) flux ratio directly and compare it to the theoretical flux for F(\10830)/F(P$\gamma$) in our $\chi^2$ analysis, rather than scaling it.  This approach is more consistent, and it incorporates the IR LUCI spectrum as its own observation on equal footing with the optical MODS spectrum.  
\beq
\frac{F(\lambda)}{F(P\gamma)} = y^{+}\frac{E(\lambda)}{E(P\gamma)}{\frac{W(P\gamma)+a_{P}(P\gamma)}{W(P\gamma)} \over \frac{W(\lambda)+a_{He}(\lambda)}{W(\lambda)}}{f_{\tau}(\lambda)}\frac{1}{1+\frac{C}{R}(P\gamma)}10^{-(f(\lambda)-f(P\gamma))C(H\beta)}  
\label{eq:F_10830_EW}
\eeq
If we were to include them, the IR Paschen lines P$\beta$ and P$\delta$ are calculated relative to P$\gamma$, and are corrected for underlying absorption based on $a_P$.  
\beq
\frac{F(\lambda)}{F(P\gamma)} = \frac{E(\lambda)}{E(P\gamma)}{\frac{W(P\gamma)+a_{P}(P\gamma)}{W(P\gamma)} \over \frac{W(\lambda)+a_{P}(\lambda)}{W(\lambda)}}\frac{1+\frac{C}{R}(\lambda)}{1+\frac{C}{R}(P\gamma)}10^{-(f(\lambda)-f(P\gamma))C(H\beta)}
\label{eq:F_IR_EW}
\eeq
Near-IR Paschen lines (P8, P9, P10, P11, P12), which are observed in the optical MODS spectrum, are calculated relative to H$\beta$.  These NIR Paschen lines are corrected for underlying absorption based on $a_P$, and H$\beta$ is corrected for underlying absorption based on $a_H$.  
\beq
\frac{F(\lambda)}{F(H\beta)}= \frac{E(\lambda)}{E(H\beta)}{\frac{W(H\beta)+a_{H}(H\beta)}{W(H\beta)} \over \frac{W(\lambda)+a_{P}(\lambda)}{W(\lambda)}}\frac{1+\frac{C}{R}(\lambda)}{1+\frac{C}{R}(H\beta)}10^{-f(\lambda)C(H\beta)}
\label{eq:F_NIR_EW}
\eeq
Finally, the blended emission line \3889 + H8 is calculated relative to H$\beta$ by summing its two fluxes contributions, each calculated according to eqs.\:(\ref{eq:F_He_EW}) \& (\ref{eq:F_H_EW}) above, respectively, except that because H8 is only separated from \3889 by 0.4 \AA, and their Doppler broadening is approximately $\sim$0.4 \AA, H8 also needs to be corrected for radiative transfer, $f_\tau(\lambda)$, just as \3889 is.  
\beq
\begin{split}
\frac{F(H\!e3889\!+\!H8)}{F(H\beta)} = & \; y^{+}\frac{E(H\!e3889)}{E(H\beta)}{\frac{W(H\beta)+a_{H}(H\beta)}{W(H\beta)} \over \frac{W(H\!e3889)+a_{He}(H\!e3889)}{W(H\!e3889)}}{f_{\tau}(H\!e3889)}\frac{1}{1+\frac{C}{R}(H\beta)}10^{-f(H\!e3889)C(H\beta)} \\
                                 & + \frac{E(H8)}{E(H\beta)}{\frac{W(H\beta)+a_{H}(H\beta)}{W(H\beta)} \over \frac{W(H8)+a_{H}(H8)}{W(H8)}}{f_{\tau}(H\!e3889)}\frac{1+\frac{C}{R}(H8)}{1+\frac{C}{R}(H\beta)}10^{-f(H8)C(H\beta)}
\label{eq:F_3889H8_EW}
\end{split}
\eeq

The model fluxes also rely on the measured equivalent widths ($W(\lambda)$).  However, the flux of the continuum at each wavelength, $h(\lambda)$, which relates the line flux to the equivalent width, is constrained such that changes in the equivalent width and changes in the emission line flux are proportional:  
\beq
F(\lambda) = W(\lambda)h(\lambda),
\eeq
where $h(\lambda)$ is fixed.  As a result, if the model parameters generate a lower flux, the equivalent width should be lowered correspondingly.  The fixed ratio $\frac{h(\lambda)}{h(H\beta)}$ is determined from the measured flux ratio and measured equivalent widths,
\beq
\frac{h(\lambda)}{h(H\beta)} = {\frac{F(\lambda)}{F(H\beta)}}_{meas} \frac{W(H\beta)_{meas}}{W(\lambda)_{meas}},
\label{eq:h_ratio}
\eeq
and the calculated flux ratios based on the model parameters are related to their corresponding equivalent widths by that fixed ratio, $\frac{h(\lambda)}{h(H\beta)}$, 
\beq
\frac{F(\lambda)}{F(H\beta)} = \frac{W(\lambda)}{W(H\beta)}\frac{h(\lambda)}{h(H\beta)}.
\eeq
Therefore, eqs.\:(\ref{eq:F_He_EW}-\ref{eq:F_3889H8_EW}) above can be rewritten to remove $W(\lambda)$ entirely and solve for a consistent emission line ratio relative to H($\beta$).  This yields a simplified set of equations for the flux ratios:  \\
Helium (Optical):  
\beq
\frac{F(\lambda)}{F(H\beta)} = y^{+}\frac{E(\lambda)}{E(H\beta)}{f_{\tau}(\lambda)}\frac{W(H\beta)+a_{H}(H\beta)}{W(H\beta)}\frac{1}{1+\frac{C}{R}(H\beta)}10^{-f(\lambda)C(H\beta)}-\frac{a_{He}(\lambda)}{W(H\beta)}\frac{h(\lambda)}{h(H\beta)}
\label{eq:F_He}
\eeq
Balmer:  
\beq
\frac{F(\lambda)}{F(H\beta)} = \frac{E(\lambda)}{E(H\beta)}\frac{W(H\beta)+a_{H}(H\beta)}{W(H\beta)}\frac{1+\frac{C}{R}(\lambda)}{1+\frac{C}{R}(H\beta)}10^{-f(\lambda)C(H\beta)}-\frac{a_{H}(\lambda)}{W(H\beta)}\frac{h(\lambda)}{h(H\beta)}
\label{eq:F_H}
\eeq 
\10830 (IR):  
\beq
\frac{F(\lambda)}{F(P\gamma)} = y^{+}\frac{E(\lambda)}{E(P\gamma)}{f_{\tau}(\lambda)}\frac{W(P\gamma)+a_{P}(P\gamma)}{W(P\gamma)}\frac{1}{1+\frac{C}{R}(P\gamma)}10^{-(f(\lambda)-f(P\gamma))C(H\beta)}-\frac{a_{He}(\lambda)}{W(P\gamma)}\frac{h(\lambda)}{h(P\gamma)}
\label{eq:F_10830}
\eeq
Paschen IR:  
\beq
\frac{F(\lambda)}{F(P\gamma)} = \frac{E(\lambda)}{E(P\gamma)}\frac{W(P\gamma)+a_{P}(P\gamma)}{W(P\gamma)}\frac{1+\frac{C}{R}(\lambda)}{1+\frac{C}{R}(P\gamma)}10^{-(f(\lambda)-f(P\gamma))C(H\beta)}-\frac{a_{P}(\lambda)}{W(P\gamma)}\frac{h(\lambda)}{h(P\gamma)}
\label{eq:F_IR}
\eeq 
Paschen Optical/NIR:  
\beq
\frac{F(\lambda)}{F(H\beta)} = \frac{E(\lambda)}{E(H\beta)}\frac{W(H\beta)+a_{H}(H\beta)}{W(H\beta)}\frac{1+\frac{C}{R}(\lambda)}{1+\frac{C}{R}(H\beta)}10^{-f(\lambda)C(H\beta)}-\frac{a_{P}(\lambda)}{W(H\beta)}\frac{h(\lambda)}{h(H\beta)}
\label{eq:F_NIR}
\eeq 
\3889 + H8:  
\beq
\begin{split}
\frac{F(H\!e3889\!+\!H8)}{F(H\beta)} = & \; (y^{+}\frac{E(H\!e3889)}{E(H\beta)}+\frac{E(H8)}{E(H\beta)}(1+\frac{C}{R}(H8)))\, \times \\ 
                                 & \; {f_{\tau}(H\!e3889)}\frac{W(H\beta)+a_{H}(H\beta)}{W(H\beta)}\frac{1}{1+\frac{C}{R}(H\beta)}10^{-f(H\!e3889+H8)C(H\beta)} \\ 
                                 & \; -\frac{(a_{He}(H\!e3889)+a_{H}(H8))}{W(H\beta)}\frac{h(H\!e3889\!+\!H8)}{h(H\beta)}
\label{eq:F_3889H8}
\end{split}
\eeq

These six equations give the updated form of eqs.\:(\ref{eq:F_He_EW}-\ref{eq:F_3889H8_EW}) above (and are presented in the same order).  For each, $\frac{h(\lambda)}{h(H\beta)}$ is defined for each flux ratio as given in eq.\:(\ref{eq:h_ratio}).  
The last of these equations, eq.\:(\ref{eq:F_3889H8}) for $F(H\!e3889\!+\!H8)$, has been algebraically simplified based on shared terms and since the continuum heights are the same for He~I $\lambda$3889 and H8.  Note also that this equation is equivalent to correcting the summed line for the sum of the underlying absorption contributions, $a_{He}$ + $a_H$, relative to the summed line's equivalent width.  

Because it is the ratio of fluxes relative to H$\beta$ and P$\gamma$ that are calculated, $W(H\beta)$ and $W(P\gamma)$ (equivalent to the H$\beta$ and P$\gamma$ flux measurements) cannot be removed from the equations above.  To treat this, $W(H\beta)$ and $W(P\gamma)$ are encoded as nuisance parameters.  For each MCMC point, a new Gaussian distributed equivalent width is chosen for each, based upon the measured value and error.  Corresponding terms are then also added into the $\chi^{2}$,
\beq
\chi^2_{W\!(H\!\beta)} = {(W(H\beta) - W(H\beta)_{meas})^2 \over \sigma^2\!(H\beta)}.
\eeq
\beq
\chi^2_{W\!(P\gamma)} = {(W(P\gamma) - W(P\gamma)_{meas})^2 \over \sigma^2\!(P\gamma)}.
\eeq
As a final point in this discussion of the relationship between the flux and equivalent width of the emission lines, it should be noted that the uncertainty on the flux, which scales the $\chi^{2}$, incorporates the uncertainty of the continuum fitting (i.e. the uncertainty of h($\lambda$)).

Finally, the temperature derived from the [O~III] $\lambda\lambda$4363, 4959, and 5007 emission lines, T(O~III), is incorporated as a prior with the resulting term added into the $\chi^{2}$:
\beq
\chi^2_{T}=(T-T_{OIII})^2/\sigma^2(T_{OIII}),
\eeq
where $\sigma(T_{OIII})=0.2~T_{OIII}$ provides a very weak constraint, but is useful in eliminating unphysical solutions (see AOS2 \citep{AOS2}).

\section{Model Data \& Sources} \label{App:Data}

\subsection{Emissivities} \label{App:Emissivities}

The helium emissivities relative to H$\beta$, $\frac{E(\lambda)}{E(H\!\beta)}$, including the collisional correction, are adopted from the work of \citeauthor{pfsd} \citep{pfsd, pfsdc}, as was implemented in AOPS \citep{AOPS}, including the use of a finer parametric grid than reported in \citeauthor{pfsd} \citep{pfsd, pfsdc}.  The helium emissivities are functions of the electron temperature and density, $T$ \& $n_e$, respectively.  The temperatures for the finer grid are in 250 K increments from 10,000 K to 25,000 K (inclusive), and 31 electron densities are irregularly spaced from 1 to 10,000 cm$^{-3}$ (also inclusive), with tighter sampling at the lower densities most relevant for helium abundance analysis.  This finer grid is available with the online version or upon request from the authors.  The emissivities were calculated for the Case B approximation with the spectral simulation code CLOUDY \citep{fer13}.  

As reported in AOS \citep{AOS}, the emissivity for H$\beta$ was provided by R.L. Porter (private communication) as a fit to the same functional form as in \citet{pfm07}, 
\beq
\begin{split}
E(H\!\beta) = & \;
[-2.6584\times10^5 + 3.5546\times10^4~\ln T - 1.4209\times10^3~(\ln T)^2 + \frac{6.5669\times10^5}{\ln T}]~\frac{1}{T} \\
              & \times 10^{-25} \textrm{ ergs cm}^{3} \textrm{ s}^{-1}.
\label{eq:HB}
\end{split}
\eeq

The helium emissivity ratio, $\frac{E(\lambda)}{E(H\!\beta)}$, is then found by interpolation on the finer parametric grid.  Measured and theoretical fluxes for \10830 are calculated relative to P$\gamma$.  Correspondingly, the emissivity ratio for \10830, $\frac{E(\lambda)}{E(P\!\gamma)}$, is calculated using the P$\gamma$ emissivity, instead of the H$\beta$ emissivity, and then interpolated on the finer parametric grid.  The P$\gamma$ emissivity is calculated from a fit to the hydrogen emissivity results of \citet{hs87},   
\beq
\begin{split}
E(P\!\gamma) = & \;
[1.0690\times10^5 - 3.1205\times10^3~\ln T - 1.4590\times10^2~(\ln T)^2 - \frac{4.8746\times10^5}{\ln T}]~\frac{1}{T} \\
               & \times[0.86877 - 2.5653\times10^{-9}~(\log_{10} n_e)^2] \times [T-10^4] \times 10^{-26} \textrm{ ergs cm}^{3} \textrm{ s}^{-1}.
\label{eq:PG}
\end{split}
\eeq

Similarly, the hydrogen emissivity ratios are calculated based on fits to the results reported by \citet{hs87}.   For all the Balmer lines and for the Optical/NIR Paschen lines in the LBT/MODS spectrum (P8, P9, P10, P11, P12), the emissivity ratio is calculated relative to H$\beta$, $\frac{E(\lambda)}{E(H\!\beta)}$.  For the IR Paschen lines from the LBT/LUCI spectrum (P$\beta$, P$\gamma$, P$\delta$), the emissivity ratio is calculated relative to P$\gamma$, $\frac{E(\lambda)}{E(P\gamma)}$.  The functional form of the fits is as follows, with table \ref{table:HEmiss} listing the coefficients, and using the definition $T_{4} = T/10^{4}$, 
\beq
\frac{E(\lambda)}{E(H\!\beta|P\gamma)} = \sum_{ij} c_{ij} (\log(T_4))^{i} (\log(n_e))^{j}.  
\eeq

\begin{table}[ht!]
\small
\centering
\vskip .1in
\begin{tabular}{lcccc}
\hline\hline
\!\!Line                           & 
{i$\downarrow$}			&
{j$\rightarrow$}  & & \\
\hline
& & 0 & 1 & 2 \\
\hline
H$\alpha$	&	0	&	2.87	&	-1.21E-3	&	-1.35E-5	\\
		&	1	&      -0.503	&	 3.11E-3	&	 7.53E-5	\\
		&	2	&	0.321	&	-2.21E-3	&	-1.02E-4	\\
H$\gamma$	&	0	&	0.468	&	 2.52E-5	&	 1.93E-6	\\
		&	1	&	2.84E-2	&	 5.25E-5	&	-1.24E-5	\\
		&	2	&      -1.53E-2	&	-4.28E-4	&	 2.63E-5	\\
H$\delta$	&	0	&	0.259	&	 3.21E-5	&	 6.09E-7	\\
		&	1	&	2.17E-2	&	-7.78E-5	&	-2.71E-6	\\
		&	2	&      -1.26E-2	&	-1.41E-4	&	 1.08E-5	\\
H8		&	0	&	0.105	&	 9.16E-6	&	 7.20E-7	\\
		&	1	&	9.42E-3	&	-3.81E-5	&	-2.16E-6	\\
		&	2	&      -6.44E-3	&	 3.41E-5	&	 1.62E-7	\\
H9		&	0	&	7.31E-2	&	 2.33E-6	&	 9.18E-7	\\
		&	1	&	6.28E-3	&	-4.33E-5	&	-1.07E-6	\\
		&	2	&      -4.58E-3	&	 4.33E-5	&	-6.12E-7	\\
H10		&	0	&	5.30E-2	&	-6.45E-6	&	 1.40E-6	\\
		&	1	&	4.27E-3	&	-3.00E-5	&	-1.47E-6	\\
		&	2	&      -3.26E-3	&	 2.51E-5	&	-2.88E-8	\\
H11		&	0	&	3.98E-2	&	-1.58E-5	&	 1.94E-6	\\
		&	1	&	2.84E-3	&	 6.17E-6	&	-3.09E-6	\\
		&	2	&      -2.34E-3	&	 7.94E-6	&	 1.06E-6	\\
H12	    &	0	&	3.06E-2	&	-2.36E-5	&	2.56E-6	\\
	    &	1	&	1.95E-3	&	1.76E-5	&	-3.77E-6	\\
	    &	2	&	-1.69E-3	&	-1.75E-5	&	2.47E-6	\\
P$\beta$	&	0	&	1.81	&	-1.99E-4	&	-1.45E-5	\\
		&	1	&      -0.171	&	 6.31E-4	&	 3.74E-5	\\
		&	2	&	8.72E-2	&	-1.09E-3	&	-1.58E-5	\\
P$\delta$	&	0	&	0.613	&	 1.11E-4	&	-3.65E-7	\\
		&	1	&	2.99E-2	&	-4.57E-4	&	 9.97E-6	\\
		&	2	&      -1.43E-2	&	 4.42E-4	&	-1.53E-5	\\
P8		&	0	&	3.65E-2	&	 1.05E-7	&	-1.37E-7	\\
		&	1	&      -8.49E-3	&	-1.12E-6	&	 2.17E-7	\\
		&	2	&      -7.94E-4	&	 5.42E-6	&	-5.27E-7	\\
P9		&	0	&	2.54E-2	&	-9.35E-7	&	 3.59E-8	\\
		&	1	&      -5.57E-3	&	-2.98E-6	&	-8.83E-9	\\
		&	2	&      -8.48E-4	&	 1.32E-5	&	-5.31E-7	\\
P10		&	0	&	1.84E-2	&	 1.35E-7	&	 4.87E-8	\\
		&	1	&      -3.93E-3	&	 2.64E-6	&	-4.63E-7	\\
		&	2	&      -6.17E-4	&	-1.11E-5	&	 8.00E-7	\\
P11	    &	0	&	1.38E-2	&	-1.49E-6	&	1.90E-7	\\
	    &	1	&	-2.91E-3	&	3.21E-6	&	-5.95E-7	\\
	    &	2	&	-5.08E-4	&	-3.27E-6	&	4.46E-7	\\
P12	    &	0	&	1.06E-2	&	-3.35E-6	&	3.27E-7	\\
	    &	1	&	-2.24E-3	&	5.54E-6	&	-8.52E-7	\\
	    &	2	&	-4.25E-4	&	1.92E-6	&	3.64E-7	\\
\hline
\end{tabular}
\caption{Fit coefficients for the hydrogen emissivities, $c_{ij}$}
\label{table:HEmiss}
\end{table}

\subsection{Reddening \& Radiative Transfer} \label{App:RRT}

The reddening coefficients, $f(\lambda)$, found in the terms $10^{-f(\lambda)C(H\beta)}$, are calculated from the extinction fits of \citet{ccm89} using $R=3.1$ ($R=\frac{A(V)}{E(B-V)}$).

The radiative transfer equations, $f_\tau(\lambda)$, in terms of optical depth, $\tau$, come from the work of \citet{bss02}.  That work does not include a fitting formula for \10830, however.  Instead, the 10-level numerical calculation program was graciously provided by Bob Benjamin (private communication), and the same functional form was fit to the \10830 optical depth data and using the same definition $T_{4} = T/10^{4}$ (see AOS4 \citep{AOS4} for details).  The added helium emission lines, He~I~$\lambda\lambda$4922 \& 5015, are singlets and are not sensitive to radiative transfer effects.  

We list here, the set of radiative transfer functions used:
\begin{align}
f_{\tau}(3889) & = 1 + (\tau/2) [-0.106+(5.14\times 10^{-5}-4.20\times 10^{-7}n_{e}+1.97 \times 10^{-10}{n_{e}}^2)T_4)] \nonumber \\
f_{\tau}(4026) & = 1 + (\tau/2) [0.00143+(4.05\times 10^{-4}+3.63\times 10^{-8}n_{e})T_4)] \nonumber\\
f_{\tau}(4471) & = 1 + (\tau/2) [0.00274+(8.81\times 10^{-4}-1.21\times 10^{-6}n_{e})T_4)] \nonumber \\
f_{\tau}(4922) & = 1  \nonumber \\
f_{\tau}(5015) & = 1  \nonumber \\
f_{\tau}(5876) & = 1 + (\tau/2) [0.00470+(2.23 \times 10^{-3}-2.51\times 10^{-6}n_{e})T_4)] \nonumber \\
f_{\tau}(6678) & = 1  \nonumber \\
f_{\tau}(7065) & = 1 + (\tau/2) [0.359+(-3.46 \times 10^{-2}-1.84\times 10^{-4}n_{e}+3.039\times 10^{-7}{n_{e}}^2)T_4)]  \nonumber \\
f_{\tau}(10830) & = 1 + (\tau/2) [0.0149+(4.45 \times 10^{-3}-6.34\times 10^{-5}n_{e}+9.20 \times 10^{-8}{n_{e}}^2)T_4)].
\label{eq:f_tau}
\end{align}

\subsection{Underlying Stellar Absorption} \label{App:UA}

The wavelength-dependent underlying absorption in terms of equivalent width for the Balmer lines is scaled relative to H$\beta$.  The Paschen lines are scaled relative to P$\gamma$, and He~I $\lambda$4471 is used for the helium lines.  For a given line, its scaling coefficient multiplies the corresponding physical model parameter---a$_{H}$, a$_{P}$, or a$_{He}$---in determining that line's underlying absorption correction.  

The adopted underlying absorption scaling coefficients are calculated from BPASS \citep{eld09, eld17, eld18}.  Our calculations are based on the BPASS spectral energy distribution output for an IMF with slope of -1.30 from 0.1 to 0.5M\sun, and a slope of -2.35 from 0.5 to 100M\sun, and z=0.001.  Both BPASS's stellar population model which includes binaries, as well as its output for a population with single stars only, were utilized.  From those two population models, absorption line equivalent widths were measured for the  hydrogen (Balmer and Paschen) and helium emission lines employed in our model.  These absorption line equivalent widths were measured for ages of 1, 2, 3 (3.16), 4, and 5 Myr.  

For each absorption line, its value was scaled to its relevant reference line:  He $\lambda$4471 for the helium lines, H$\beta$ for the Balmer lines, and P$\gamma$ for the Paschen lines.  These ratios are relatively constant over the ages investigated, though the width of the strong hydrogen absorption lines for the earliest ages makes setting the continuum properly challenging for the higher, more closely spaced, lines, and the metallicity evolution of the continuum makes measuring the weaker helium more challenging at the later ages.  Correspondingly, clear outliers were excluded, and this primarily affected the higher Balmer and Paschen lines at 1 \& 2 Myr.  

Since the underlying absorption ratios were relatively constant over age and the single-only and binaries-included populations, the ratio values for each line were averaged to determine the scaling coefficients for our model.  For the Paschen lines, the ratios were all very close to unity.  The deviations from unity in the averages between the Paschen lines were not established to be significant and not instead due dominantly to random variation.  As a result, the underlying absorption correction (in terms of equivalent of width) for all of the Paschen lines was taken to be the same, with no variation with wavelength.  Because it is blended with H8, the absorption profile for \3889 could not be directly measured.  Correspondingly, for \3889 we continue to use the value calculated by F. Rosales-Ortega (private communication) and employed in our previous work.  See AOS \citep{AOS} for details.  All other employed underlying absorption scaling coefficients are calculated from our measurements based on the BPASS spectra.  

The adopted scaling coefficients are given in table \ref{table:UAcoeffs}.  

\begin{table}[ht!]
\centering
\vskip .1in
\begin{tabular}{llllllllll}
\hline\hline
\!\!Balmer	&	H$\alpha$	&	H$\beta$	&	H$\gamma$	&	H$\delta$	&	H8	&	H9	&	H10	&	H11	&   H12		\\
	&	0.930	&	1	&	0.964	&	0.927	&	0.893	&	0.868	&	0.824	&	0.778	&	0.750	\\
\hline
Paschen	&		&	P$\beta$	&	P$\gamma$	&	P$\delta$	&	P8	&	P9	&	P10	&   P11 &   P12		\\
	&		&	1	&	1	&	1	&	1	&	1	&	1	&	1	&	1	\\
\hline
He~I~$\lambda\lambda$:  	&	3889	&	4026	&	4471	&	4922	&	5015	&	5876	&	6678	&	7065	&	10830	\\
	&	1.4	&	1.156	&	1	&	0.632	&	0.420	&	0.708	&	0.461	&	0.346	&	0.750	\\
\hline
\end{tabular}
\caption{Underlying Absorption Scaling Coefficients by Line}
\label{table:UAcoeffs}
\end{table}

\subsection{Neutral Hydrogen Collisional Emission} \label{App:NHCC}

To account for the contribution to the hydrogen fluxes from neutral hydrogen collisional excitation, the relative amount of collisional-to-recombination emission can be calculated as follows,
\beq
\frac{C}{R}(\lambda) = \frac{n(H~I)(\sum_{i}K_{1 \rightarrow i}BR_{i \rightarrow j})BR_{j \rightarrow 2} n_{e}}{n(H~II) \alpha_{+ \rightarrow j}BR_{j \rightarrow 2} n_{e}} = \frac{n(H~I)}{n(H~II)}\frac{K_{eff}}{\alpha_{eff}} = \xi \frac{K_{eff}}{\alpha_{eff}}.
\label{eq:CRsum}
\eeq
In eq.\:\ref{eq:CRsum}, \textit{K} represents the collisional transition rate from the ground state to level \textit{i} above the transition level of interest, \textit{j} ($j \rightarrow 2 = \lambda$).  Downward transitions from \textit{i} occur with a variety of branching paths.  Correspondingly, \textit{$BR_{i \rightarrow j}$} represents the relevant branching fraction for ultimately transitioning to \textit{j}.  \textit{$\alpha$} is the effective recombination rate to level \textit{j} (\textit{$\alpha_{+ \rightarrow j}=\alpha_{eff}$} with ``$+ \rightarrow j$'' serving as an analogue to \textit{$K_{1 \rightarrow i}$} but for ionization recombining to level \textit{j}).  \textit{$K_{eff}$} is the sum over the excitation levels of the collisional transition rate, \textit{$K_{1 \rightarrow i}$}, times the corresponding branching fraction, \textit{$BR_{i \rightarrow j}$}.  Given the relative complexity of the definition of \textit{$\frac{C}{R}$}, it is useful to note that the electron density and branching fraction from level \textit{j} to level 2 cancel (or, similarly for Paschen lines, the corresponding branching fraction from level \textit{j} to 3 cancels), and we are left with the ratio of the neutral hydrogen density, n(H~I), to the ionized hydrogen density, n(H~II), times a collisional rate over a recombination rate.  The ratio $\frac{n(H~I)}{n(H~II)}$ is then defined as the model parameter $\xi$.  A typical value of $\xi$ for a H~II region is $\sim 10^{-4}$.  Correspondingly, we define $\xi_{4} = \xi \times 10^{4}$, such that,   
\beq
\frac{C}{R}(\lambda) = {\xi_{4}} \times 10^{-4} \times \frac{K_{eff}}{\alpha_{eff}}.
\label{eq:CR}
\eeq

As defined above, \textit{$\frac{C}{R}$} makes the simplifying assumption that all of the neutral hydrogen is excited from the ground state.  At the temperatures and densities relevant for the H~II regions employed in helium abundance analysis, this is a reasonable assumption.  In AOS \citep{AOS}, the collisional rates were based on power law fits ($A T^{B}$) to the effective collision strengths, $\Upsilon$, reported in \citet{and02} up to a principle quantum number of $n = 5$.  C.P. Ballance (private communication) graciously extended those calculations to $n = 8$.  The approach follows AOS \citep{AOS} but with the updated collisional data allowing for more accurate modeling of the collisional excitation and the inclusion of emission lines resulting from higher excitations.  
\begin{eqnarray}
K_{i \rightarrow j}=\frac{2\sqrt{\pi}\alpha ca_{0}^{2}}{\omega_{i}}\sqrt{\frac{I_{H}}{k_{B}T}}\exp(-\frac{\triangle E_{ij}}{k_{b}T})\Upsilon_{ij}, \\
K_{1 \rightarrow i}=4.004\times10^{-8}\sqrt{\frac{1}{k_{B}T}}\exp(\frac{-13.6(1-\frac{1}{i^{2}})}{k_{b}T})\Upsilon_{1i}, \nonumber
\end{eqnarray}
with $I_{H}=-13.6~eV$ the ionization potential, $a_0$ is the Bohr radius, and $\omega_{i}=(2s+1)(2l+1)$ is the statistical weight of the level.
The branching fractions are calculated directly from the Einstein transition coefficients as reported in \citet{om83}, and the effective recombination rate is fit from the case B data of \citet{hs87} for an electron density of 100 per cm$^{3}$ and again employing a power law fit.  In theory, the collisional sum includes an infinite number of levels, but the probabilities fall off rather quickly.  For this work the sum excludes any terms contributing less than 1\%.  

\indent For H$\alpha$, 3s, 3p, 3d, 4s, 4d, \& 4f are included;\\
\indent For H$\beta$, 4s, 4p, 4d, 4f, 5s, 5d, \& 5f;\\
\indent For H$\gamma$$|$P$\beta$, 5s, 5p, 5d, 5f, 5g, 6s, \& 6f;\\
\indent For H$\delta$$|$P$\gamma$, 6s, 6p, 6d, 6f, 6g, 6h, \& 7s;\\
\indent For P$\delta$, 7s, 7p, 7d, 7f, 7g, 7h, \& 7i;\\
\indent For H8$|$P8, 8s, 8p, 8d, 8f, 8g, 8h, 8i, \& 8j;\\

For H9, H10, H11, H12, P9, P10, P11, \& P12, collisional excitations to $n$ = 9, 10, 11, \& 12 are required.  As discussed in \citet{fmdzb16}, the effective collisional strength scales inversely proportional to the principle quantum number cubed, 
\beq
\Upsilon(n) \sim \frac{A}{(n+\alpha)^3}. 
\eeq
To estimate the collisional excitation for H9 \& P9, H10 \& P10, H11 \& P11, and H12 \& P12, the $\ell$ contributions for each level up to $n=8$ were summed and the $1/n^3$ scaling law above fit to $n$ = 5, 6, \& 7 and used to extrapolate to $n$ = 9, 10, 11, \& 12.  An effective branching fraction was calculated by extrapolating the collisionally weighted average.  

Table \ref{table:HIcoll} lists the coefficients for constructing $\frac{K_{eff}}{\alpha_{eff}} \times 10^{-4}$, the quantity that multiplies $\xi_{4}$ and provides the increasing collisional enhancement factor for increasing temperature ($T_{4} = \frac{T}{10^{4}}$),
\beq
\frac{C}{R}(\lambda) = {\xi_{4}} \times 10^{-4} \times \frac{K_{eff}}{\alpha_{eff}} = {\xi_{4}} \times \sum_{i}a_{i}\exp(-\frac{b_{i}}{T_{4}}){T_{4}}^{c_{i}}.
\label{eq:CR2}
\eeq

\begin{figure}
\resizebox{\textwidth}{!}{\includegraphics{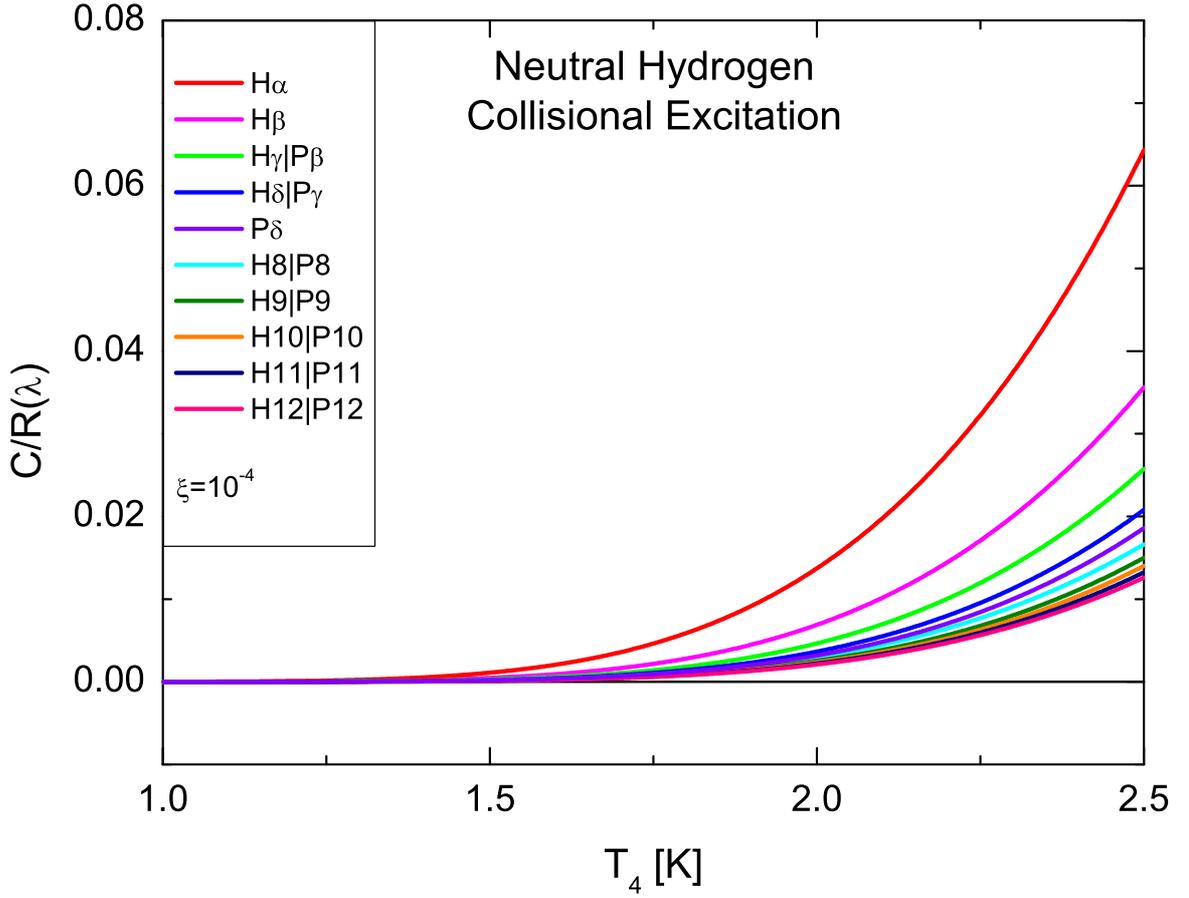}}
\caption{
The collisional excitation relative to recombination rate for the hydrogen emission lines due to neutral hydrogen.  The behavior is dominantly exponential with temperature, with, as expected, larger corrections for the lower level (lower energy) emission lines.  Note that the enhancement will scale linearly with increasing the neutral to ionized hydrogen ratio, $\xi$ (a characteristic value of 10$^{-4}$ is used for the plot).
}
\label{H_CR_T}
\end{figure}

Figure~\ref{H_CR_T} shows the exponential temperature dependence of $\frac{C}{R}$ and compares the relative collision contributions by emission line.  

\begin{table}[ht!]
\small
\footnotesize
\centering
\vskip .1in
\begin{tabular}{lccccccccc}
\hline\hline
Line	&	Coefficient	&	Terms	&	&	&	&	&	&	& \\
\hline
H$\alpha$	&	a	&	0.2914	&	2.5037	&	2.2940	&	0.2761	&	4.5521	&	0.3542	&		&		\\
	&	b	&	-14.796	&	-14.029	&	-14.029	&	-14.796	&	-14.029	&	-14.796	&		&		\\
	&	c	&	0.3544	&	0.5765	&	0.5776	&	0.6016	&	0.6585	&	0.6705	&		&		\\
\hline
H$\beta$	&	a	&	0.4885	&	0.1321	&	1.8224	&	1.4275	&	0.0971	&	2.7755	&	0.1444	&		\\
	&	b	&	-14.796	&	-15.151	&	-14.796	&	-14.796	&	-15.151	&	-14.796	&	-15.151	&		\\
	&	c	&	0.4315	&	0.4999	&	0.6788	&	0.7476	&	0.7777	&	0.7793	&	0.8645	&		\\
\hline
H$\gamma$	&	a	&	0.1310	&	0.5530	&	0.0606	&	1.3800	&	0.9666	&	0.0756	&	1.8195	&		\\
	&	b	&	-15.151	&	-15.151	&	-15.344	&	-15.151	&	-15.151	&	-15.344	&	-15.151	&		\\
	&	c	&	0.2098	&	0.5533	&	0.6625	&	0.8310	&	0.9178	&	0.9553	&	0.9861	&		\\
\hline
H$\delta$	&	a	&	0.0665	&	0.1686	&	0.4932	&	1.1081	&	1.5477	&	0.7519	&	0.0463	&		\\
	&	b	&	-15.344	&	-15.344	&	-15.344	&	-15.344	&	-15.344	&	-15.344	&	-15.460	&		\\
	&	c	&	0.0857	&	0.3288	&	0.7005	&	0.9260	&	0.9872	&	0.9933	&	1.0062	&		\\
\hline
H8	&	a	&	0.0596	&	0.0986	&	0.1369	&	0.2691	&	0.5996	&	0.6479	&	1.1273	&	1.3033	\\
	&	b	&	-15.536	&	-15.536	&	-15.536	&	-15.536	&	-15.536	&	-15.536	&	-15.536	&	-15.536	\\
	&	c	&	-0.1141	&	0.1183	&	0.2193	&	0.4039	&	0.6536	&	0.7722	&	0.8048	&	0.8441	\\
\hline
H9	&	a	&	3.1377	&		&		&		&		&		&		&		\\
	&	b	&	-15.587	&		&		&		&		&		&		&		\\
	&	c	&	0.9733	&		&		&		&		&		&		&		\\
\hline
H10	&	a	&	2.9332	&		&		&		&		&		&		&		\\
	&	b	&	-15.624	&		&		&		&		&		&		&		\\
	&	c	&	0.9890	&		&		&		&		&		&		&		\\
\hline
H11	&	a	&	2.7673	&		&		&		&		&		&		&		\\
	&	b	&	-15.652	&		&		&		&		&		&		&		\\
	&	c	&	1.0021	&		&		&		&		&		&		&		\\
\hline
H12	&	a	&	2.6300	&		&		&		&		&		&		&		\\
	&	b	&	-15.672	&		&		&		&		&		&		&		\\
	&	c	&	1.0130	&		&		&		&		&		&		&		\\
\hline
P$\beta$	&	a	&	0.1310	&	0.5530	&	0.0606	&	1.3800	&	0.9666	&	0.0756	&	1.8195	&		\\
	&	b	&	-15.151	&	-15.151	&	-15.344	&	-15.151	&	-15.151	&	-15.344	&	-15.151	&		\\
	&	c	&	0.2098	&	0.5533	&	0.6625	&	0.8310	&	0.9178	&	0.9553	&	0.9861	&		\\
\hline
P$\gamma$	&	a	&	0.0665	&	0.1686	&	0.4932	&	1.1081	&	1.5477	&	0.7519	&	0.0463	&		\\
	&	b	&	-15.344	&	-15.344	&	-15.344	&	-15.344	&	-15.344	&	-15.344	&	-15.460	&		\\
	&	c	&	0.0857	&	0.3288	&	0.7005	&	0.9260	&	0.9872	&	0.9933	&	1.0062	&		\\
\hline
P$\delta$	&	a	&	0.0792	&	0.0890	&	0.1853	&	0.5125	&	0.9682	&	0.6412	&	1.2751	&		\\
	&	b	&	-15.460	&	-15.460	&	-15.460	&	-15.460	&	-15.460	&	-15.460	&	-15.460	&		\\
	&	c	&	0.0012	&	0.1686	&	0.4715	&	0.7576	&	1.0176	&	1.0350	&	1.0717	&		\\
\hline
P8	&	a	&	0.0596	&	0.0986	&	0.1369	&	0.2691	&	0.5996	&	0.6479	&	1.1273	&	1.3033	\\
	&	b	&	-15.536	&	-15.536	&	-15.536	&	-15.536	&	-15.536	&	-15.536	&	-15.536	&	-15.536	\\
	&	c	&	-0.1141	&	0.1183	&	0.2193	&	0.4039	&	0.6536	&	0.7722	&	0.8048	&	0.8441	\\
\hline
P9	&	a	&	3.1377	&		&		&		&		&		&		&		\\
	&	b	&	-15.587	&		&		&		&		&		&		&		\\
	&	c	&	0.9733	&		&		&		&		&		&		&		\\
\hline
P10	&	a	&	2.9332	&		&		&		&		&		&		&		\\
	&	b	&	-15.624	&		&		&		&		&		&		&		\\
	&	c	&	0.9890	&		&		&		&		&		&		&		\\
\hline	
P11	&	a	&	2.7673	&		&		&		&		&		&		&		\\
	&	b	&	-15.652	&		&		&		&		&		&		&		\\
	&	c	&	1.0021	&		&		&		&		&		&		&		\\
\hline
P12	&	a	&	2.6300	&		&		&		&		&		&		&		\\
	&	b	&	-15.672	&		&		&		&		&		&		&		\\
	&	c	&	1.0130	&		&		&		&		&		&		&		\\
\hline
\end{tabular}
\caption{Coefficients for the Neutral Hydrogen Collisional Correction}
\label{table:HIcoll}
\end{table}

\end{appendices}

\end{document}